**Anomalies and fluctuations of near-surface air temperature at Tianhuangping (Zhejiang), China, produced by the longest total solar eclipse of the 21st century under cloudy skies**

Marcos A. Peñaloza-Murillo[a]

*Universidad de los Andes, Facultad de Ciencias, Departamento de Física
Mérida 5101, Edo. Mérida, Venezuela
&
Williams College, Department of Astronomy
Williamstown, Massachusetts 01267, U.S.A.*

Michael T. Roman[b]

*The University of Leicester, Department of Physics and Astronomy
Leicester LE1 7RH, U.K.*

Jay M. Pasachoff[c]

*Williams College, Department of Astronomy
Williamstown, Massachusetts 01267, U.S.A.
&
Carnegie Observatories, Pasadena, California 91101, U.S.A.*

Emails: (a) map4@williams.edu, (b) mr359@le.ac.uk, (c) eclipse@williams.edu

**Abstract**

We analyze the near-surface air temperature response, at three different heights over the ground, recorded by the Williams College expedition under meteorological conditions characterized by cloudy skies during the longest total solar eclipse of the 21st century on 22 July 2009, at Tianhuangping (Zhejiang), China. An analysis of the relationship between solar radiation and air temperature was made by applying a study previously published in which we evaluated the cloudiness contribution in estimating the impact on global solar radiation throughout this phenomenon at that site. The analysis of this response includes linear and absolute negative anomalies as well as fluctuations, which was undertaken through a statistical study to get information on the convection activity produced by the latter. The fluctuations generated by turbulence were studied by analyzing variance and residuals. The results, indicating a steady decrease and recovery of both perturbations, were consistent with those published by other studies for this total solar eclipse.

## 1. Introduction

The observation and study of near-surface air temperature (NSAT) variation during a solar eclipse has always played a relevant direct or indirect role in atmospheric science since it first began to be measured in the nineteenth century (Brereton, 1834; Alexander, 1854) [see also Table S1 in Supplementary on-line material of Peñaloza-Murillo & Pasachoff (2015) for some other historical references]. Subtle atmospheric effects should be expected and detected due to the relatively rapid change in insolation and temperature, and others ambient variables, during a total solar eclipse (TSE) [for an extensive (but not exhaustive) list of references, see Tables S2-S3 in Supplementary On-line material of Peñaloza-Murillo & Pasachoff (2015)]. The greatest thermal response and consequent effects would be expected during TSEs with greatest durations but under clear skies (Aplin & Harrison, 2003); therefore, relatively long-duration TSEs should provide the best opportunity to measure and analyze atmospheric response under this condition. Scientific observations at total solar eclipses were summarized by Pasachoff (2009, 2017).

According to Meeus (2002) and Espenak (2014) between 2001 and 2100, there are eleven TSEs with maximum duration greater than five minutes, with none more than seven minutes [see Table 1 of Peñaloza-Murillo & Pasachoff (2018) for TSEs with maximum totality duration greater than five minutes during this century]. The longest one, with up to 6' 40" of totality, took place on 22 July 2009, with its longest duration somewhere in the western Pacific Ocean. Part of its shadow path passed over India, Nepal, Bhutan, Bangladesh, China and some Japanese islands in the morning (Fig. 1). The expedition of Williams College – Hopkins Observatory to observe this event went to China (Pasachoff, 2011), where its scientific equipment was set up at Tianhuangping reservoir (Anji, Zhejiang), to be close to as far east as possible on land before the path of totality went over the ocean [see

http://web.williams.edu/astronomy/eclipse/eclipse2009/2009total/index.html].

NSAT measurements were taken there at three different heights over the ground and, in this study, they are presented and analyzed following the goals and the general procedure described and used in a previous paper (Peñaloza-Murillo & Pasachoff, 2015) in which an air-cooling analysis was made based on the air temperature observations taken during the first TSE of the 21[st] century on 21 June 2001, at Lusaka, Zambia (Africa) by the Williams College – Hopkins Observatory expedition to Africa, and in another previous paper (Peñaloza-Murillo & Pasachoff, 2018) in which cloudiness and solar radiation were also analyzed during this TSE of 2009, the longest one of the 21st century, in China.

## 2. The longest TSE of the 21[st] century at Tianhuangping (Zhejiang), China: local circumstances

### 2.1 Local geographical data and astronomical circumstances

The site selected for the observations were the grounds of the Tianhuangping pumped-storage power station located in Anji County, Zhejiang Province, near Hangzhou, about 175 km to the west of Shanghai. This station has an upper reservoir or artificial lake on the top of a mountain, surrounded by green bamboo forests. The geographical coordinates of the site are 119° 35' 28" E, 30°28' 07" N and 890 m above sea level. Table 1 gives the astronomical and geographical circumstances for this event. From these data it can be observed that we are dealing with a TSE that occurred during the morning, which has significant implications for our analysis as will be shown later.  The sunrise was at 05:12:53, first contact at 08:20:49.1, and 4[th] contact at 10:57:59.5. Contacts and other interval times are given in local time (LT = UTC - 8 h). The total phase occurred between 09:33:02.8 and 09:38:44.9. The duration of totality was ~ 5.70 min and the total duration of the eclipse was of ~ 2.62 h. The site was about 30.88 km away from the central line with a width of totality of ~ 267 km and Xavier Juvier calculated a value of 0.872 km/s for the umbra velocity at this particular place.

[http://xjubier.free.fr/en/site_pages/solar_eclipses/TSE_2009_GoogleMapFull.html?Lat=30.46861&Lng=119.59111&Elv=890.0&Zoom=5&LC=1].

When the center of the umbra reached the Indian-China border (at 01:05 TCU), the shadow's velocity was 1.8 km/s; at Hubei province's capital Wuhan, just 20 km south of central line, the umbra's velocity was 1.0 km/s (at 01:27 UTC) (Espenak & Anderson, 2008). The instant of greatest eclipse occurred at 02:35:19 UTC at 24° 13' N and 144° 07' E when axis of the Moon's shadow passed closest to the center of Earth.

*2.2 Meteorological circumstances*

At that date, in the morning and during the event, the sky was mostly cloudy over the region of Zhejiang. Morning storms were followed by widespread cirrostratus and variable stratocumulus during the eclipse, followed by afternoon rain showers. Because of its importance in assessing the incident solar radiation on the ground and, therefore in the overall eclipse, this cloudiness, recorded with a wide-field camera, is quantified and analyzed in Peñaloza-Murillo & Pasachoff (2018). Table 2 gives some meteorological circumstances (and also photometric) corresponding to which this event took place, conveying information on variation of different variables throughout and surrounding the eclipse period. Note that NSAT measurements are given for three different heights above the ground: very close to the ground at 2 cm, close to the ground at 10 cm, and above the ground at 2 m. The photometric and relative-humidity information was taken from the measurements made by Stoeva et al. (2009) at the same observation site. These authors gave illumination values for three different directions in the sky (zenith, horizon and solar area).

Total reductions of 6.01 °C, 5.90 °C, and 4.37 °C in air temperature were observed at 2 cm, 10 cm, and 2 m, respectively, in duration intervals of ~ 80 min (1.30 h), ~ 79 min (1.32 h) and ~ 79 min (1.32 h), respectively, between first contact and the instant of minimum temperature (see Table 2). This same feature was observed for the respective time lag: 7.98 min for the lower height and 7.08 min for the rest, measured from the beginning of totality (at second contact) when the direct solar radiation is first completely blocked by the Moon. Recall that thermal response of the air is not in phase with the solar radiation response. The drop in temperature during these lags were, respectively, 0.73 °C, 0.44 °C and 0.34 °C. Other reductions such as the maximum "linear anomaly" $\Delta T_{LIN}$, were 8.42 °C, 7.07 °C, and 5.27 °C, respectively (see Table 2). By applying a linear regression fit, this anomaly is related to a hypothetical temperature time series over the eclipse period, using observed temperatures at the time of first and last contacts [Ramesh et al., 1982 (see their Fig. 4, middle panel); Segal et al., 1996 (see their Fig. 3); Clark, 2016 (see Fig. 2b)]. This linear time series is assumed to be representative of temperature observations in the absence of an eclipse. Thus, the linear anomaly is obtained by subtracting the value of the interpolated temperature corresponding to that under no-eclipse condition ($T_{i,n-e}$) at the same time of that corresponding to the observed minimum temperature during the eclipse. The other anomaly, the "absolute anomaly" $\Delta T_{ABS}$, were 6.80 °C, 6.77 °C and 5.17 °C, respectively (see Table 2), which involves calculating the difference between a pre-eclipse maximum temperature (between first and second contacts) and an eclipse period minimum temperature [see Fig. 2a (Clark, 2016)].

Changes in surface temperatures moves the air's water vapor content closer to saturation, increasing its relative humidity (*RH*). This time *RH* increased by 12% between 09:10:00 and 09:45:00 (Stoeva et al., 2009), corresponding from roughly 23 minutes prior to 6 minutes following totality. The ambient illumination was drastically reduced in all three directions by up to 17 lx, for the zenith; 4 lx for the horizon; and 120 lx, for the solar area (Stoeva et al., 2009). No wind data were available.

### 3. Instrumentation and air temperature measurements

Near-ambient air temperature was measured with a MicroDAQ HOBO U12 data logger with 4 external inputs suitable for multiple parameters of outdoor environmental measurements and recording. It is capable of measuring and storing up to 43,000 12-bit samples/readings with an operating range of - 20°C to 70°C and with a resolution of 0.6 mV. It has an adjustable sampling rate of 1 second to 18 hours. More details about this equipment can be found in [https://www.microdaq.com/onset-hobo-u12-outdoor-industrial-data-logger.php]. A calibration checking of the instrument was carried out days prior to the event. The temperature sensors were installed at three different heights above the ground (see Table 2), surrounded by green vegetation, and shielded from direct sunlight using small shades. Figure 3 presents a general view of temperature variation from 06:13:00 to 14:44:00. Readings were recorded every 63 seconds. An inspection of this figure shows the effect of the eclipse on this variable in the morning after the eclipse ended as well as the effect of bad weather after midday and during the first part of the afternoon. In Fig. 4, we show only the section corresponding to NSAT variation during the eclipse from first to fourth contact, roughly from 08:21 to 10:58 at 1-minute intervals. Graphically, all three curves show a delay or lag in relation to the central phase, which has been quantified from the data (see also Table 2). The drop of temperature is more significant at 2 cm height than the other two heights. Generally, the surface layer is the warmest due its contact (and consequent thermal conduction) with the radiatively heated ground; therefore, it cools down more rapidly (see Fig. 3).

The whole pattern displayed by our measurements in Fig. 3 is typical for cloudy or/and rainy weather. It is quite similar (mainly after 11:00 up to past 14:00) to that published by Pintér et al. (2010) for the same TSE and to that published by Winkler et al. (2001) for the 1999 TSE at Garching, Germany.

### 4. Analysis of the relationship between solar radiation and air temperature

In a previous paper, Peñaloza-Murillo & Pasachoff (2015) presented detailed methodological considerations in relation to the problem of how to analyze the relationship between solar radiation and air temperature during a total solar eclipse following the works of Phillips (1969), Szałowski (2002), Tzanis (2005), and Pintér et al. (2008), though those previous works were for relatively clear skies. The reader is referred to all these papers for details. Yet this time we face an additional problem: how to analyze mathematically this relationship under significant cloudy skies as those we had at Tianhuangping (Peñaloza-Murillo & Pasachoff, 2018)

The umbra of the last TSE of the 20th century, which took place on August 11, 1999, pass over Western, Central, and Eastern Europe, under variable cloudy conditions. A considerable number of papers, however, were published on the meteorological effects of this eclipse [see Table S1, on-line supplementary material of Peñaloza-Murillo & Pasachoff (2015)] but none of them attempted to undertake a specific mathematically analysis between solar radiation and local NSAT. The worst region for observing this occultation was England, where the sky was heavily cloudy or practically overcast (e.g., Camborne, Cornwall); even so, some micro-meteorological measurements were made at several sites in that country [e.g., Hanna (2000), Leeds-Harrison et al. (2000), Morecroft et al. (2000), Aplin & Harrison (2003)]. In particular the latter authors tried to interpret theoretically their temperature measurements in the light of the diurnal-cyclones theory and the cold-core eclipse-cyclone hypothesis proposed by Clayton (1901) at the onset of the previous century [to update research on Clayton's cold-core cyclone hypothesis, see Gray & Harrison (2012)].

Our first challenge was to obtain a theoretical model of solar radiation under cloudy and eclipse conditions for our observation site from which we can derive theoretical models for NSAT measurements presented in the curves of Fig. 4. For these goals it was necessary first to get the occultation and obscuration function, via limb-darkening integration as was done by Peñaloza-Murillo & Pasachoff (2018) and to which the reader is referred for details. Our challenge now is to apply the pioneering Phillips's method (Phillips, 1969), in combination with that of Peñaloza-Murillo & Pasachoff (2018), to retrieve the temperature profile directly from a radiative model of the solar radiation's collapse and recovery during the eclipse process, this time total, but under cloudy conditions.

Phillips's method is described more thoroughly in Peñaloza-Murillo & Pasachoff (2015). It is based on a sort of calibration curve of the type "air temperature vs. solar radiation" from which we can extract the air temperature during the phenomenon. In this case, clouds of different types and heights affected the solar radiation before, during, and after the eclipse (Peñaloza-Murillo & Pasachoff, 2018). Therefore, our first task, of this part of the investigation, was to formulate a solar-radiation model under cloudy conditions, as if the occultation had not happened, in order to be taken subsequently as such in the eclipse situation, as was obtained by Peñaloza-Murillo & Pasachoff (2018) (see Fig. 5).

*4.1 NSAT profiles*

To a first approximation the daily ambient air temperature, and in particualr the NSAT, for a given location, is directly related to daily global radiation, modulated, of course, by many factors including weather and geography. The idea that this relation could be used to model air temperature during a solar eclipse under clear skies came first from the work of Phillips (1969). It has successfully been applied by Peñaloza-Murillo & Pasachoff (2015) in a cloudless situation. Phillips proposed a methodology based on the construction of a plot relating air temperature to global solar radiation, omitting temperatures measured during the occultation. Global solar radiation values were obtained from the corresponding model. In this way, a sort of calibration curve of the type "Air Temperature vs. Global Solar Radiation" was produced by interpolating a final regression straight line in the sector of the plot corresponding to the omissions of eclipse-related values of air temperature. Given the values of this radiation provided by the model during the eclipse (Fig. 4), one can retrieve air temperature values by reading out the calibration curve.

For every temperature profile of Fig. 4 we constructed calibration charts, which provided the calibration curves (Fig. 6). From these curves, the instantaneous thermal response of the air, at the three specified heights, were found as depicted in Fig. 7. Here, instantaneous thermal response stands for the air temperature that should be attained in absence of any kind of delay or lag, in phase with the solar radiation change and also with the obscuration function. Defined in this way, this variable can be called "instantaneous temperature," $T_{inst}$.

In comparing these instantaneous profiles to measurements as shown in Figs. 8 - 10, [see also Figs. 11 (a), 12 (a) and 13 (a)], some interesting features emerge. The measurements exhibit appreciable fluctuations or instabilities between 08:21 and 09:07 for +2 cm and, from 08:21 to 09:10 for both +10 cm and +2 m. The fluctuations begin to disappear as totality approaches (09:37) with little to no fluctuations from 09:08 to 10:07 at +2 cm, from 09:11 to 10:03 at +10 cm and between 09:11 and 09:48 at +2 m. The fluctuations appear again from these points until 10:57, around fourth contact. The fluctuations are stronger during the first phase of the eclipse than in the final phase. These fluctuations are interpreted as convective activity in the surface layer (to be analyzed in the next section). The reduction in convective instability is likely a direct result of the eclipse. Surprisingly, during the period of convective turbulence and consequent temperature fluctuations preceding totality (i.e. prior to ~9:16 LT), the measured temperatures appeared cooler than expected from our instantaneous radiative model; this suggests that the atmospheric cooling was enhanced by dynamical processes not captured in our radiative model (e.g. local advection, vertical mixing) or simply that our radiative model failed to completely account for the observed cloud cover during this period. This effectively results in cooling that appears to precede that seen in our instantaneous model, and clearly illustrates how clouds can alter the observed temperatures. In a cloudless sky, however, this effect does not occur, as has been shown in our previous investigation of the African TSE of 2001 (Peñaloza-Murillo & Pasachoff, 2015) and others.

In above context, it is worthy to note that Aplin & Harrison (2003) found a similar effect at Camborne (England), during the TSE of August 11, 1999, under overcast skies. The temperature minimum at that eclipse actually occurred *before* totality was reached and not after, as expected. This inverse delay can be referred to as a "negative lag" due to cloudiness. In the present case some negative lags are observed clearly in Figs. 8, 10 and to a lesser extent in Fig. 9. As long as the effect of decreasing cloudiness tended to be small (as it was observed around totality), the measurements, in a first stage and up to some point, came earlier than the theoretical values (negative lag); afterwards and inversely, the theoretical values came first, that is to say earlier, then the increasing values of the temperature measurements, as it is normally expected (positive lag). Toward the final partial phase after totality, at +2 cm (Fig. 8), we can observe another lag inversion. This temporal inter-changeability between negative and positive lag is the result of the combined effect of clouds on solar radiation variability, during the eclipse, and the natural thermal complex response of the ambient air to the phenomenon (Rabin & Doviak, 1989). See Figs. 8 – 10 for more details.

For the remainder of this work, we will focus on analysing the NSAT fluctuations. The analysis and quantification of negative lag found at the different heights n this research will be made in future work.

## 5. Analysis of air temperature fluctuations

The notable existence of NSAT fluctuations, of different magnitude, during the eclipse is evident for the three sensing situations presented in Fig. 4 (see also Figs. 8 - 10). These fluctuations are a recognizable phenomenon attributable to convective turbulence, which in general tends to disappear or minimize as long as the sunlight diminishes (Nieuwstadt & Brost, 1986) including that due to the eclipse (Kadygrov et al., 2013), and return to normal conditions following the eclipse when this event occurs in the morning. This same recovery would likely not be seen in an afternoon eclipse (Peñaloza-Murillo & Pasachoff, 2015). This convection suppression led to a more stable condition of the air in a certain period of time around totality, for the three cases considered here. The mechanism behind these changes is a product of the abrupt alteration in insolation, causing cooling in the surface layers of the atmosphere and damping of atmospheric turbulence from the surface upwards. Mixed with these fluctuations, there are those coming from cloudiness variation.

To analyze these fluctuations, as in our previous work (Peñaloza-Murillo & Pasachoff, 2015), we follow the method of Szałowski (2002) based on a variance analysis and residuals calculations of the convective turbulence. In doing so, it is convenient to make our analysis separately to each of the three cases involved in this investigation: case I, at 2 cm; case II, at 10 cm; and case III, at 2 m above the ground, respectively.

*5.1 Case I (+ 2 cm)*

Focusing on the measurements of Fig. 4 in detail (or Fig. 8), we observe in this case two temporal segments in which the convective activity is remarkable. The first, at the beginning, goes from 08:21 to 09:06. The other at the end, goes from 10:02 to 10:57. In between, we have the stable segment corresponding to the interval 09:07 to 10:02. We note, however, that the first exhibits appreciable fluctuations in the sub-interval 08:21 - 08:46 (sub-interval $A_I$) and smaller ones in the next sub-interval 08:47 to 09:06 (sub-interval $B_I$). For the last segment, we have similar behavior. In the sub-interval 10:03 - 10:35 (sub-interval $D_I$), we have small fluctuations, and in the final sub-interval, from 10:36 to 10:57 (sub-interval $E_I$), the convective activity becomes increasingly higher again in accordance to what one can expect late in the morning. The middle stable segment is designed, then, as sub-interval $C_I$. Under this temporal distribution of five sub-intervals, to some extent arbitrarily chosen on the base of an only visually different temperature fluctuation, we proceed to apply the method of Szałowski (2002) to each of these sub-intervals.

The method states that the variance is a common and good parameter to measure directly temperature fluctuations owing to convectional turbulent activity but certain conditions apply. Szałowski (2002) defines the temperature fluctuation as:

$$T'(t) = T(t) - \bar{T}(t) , \qquad (1)$$

where $\bar{T}(t)$ represents the mean temperature and $T(t)$ represents the measured temperature at time $t$. If this mean varies negligibly, as happens especially for short measurements series investigating high frequency temperature fluctuations in stable weather conditions, the analysis of variance $\sigma_{T'}$ is direct. If, on the other hand, the fluctuation process is random, it implies that the mean of fluctuations is zero, $\bar{T}' = 0$, allows for estimating $\bar{T}(t)$ by least-squares fit. Then, $\bar{T}(t_i) = T_{pred}$ and the residuals, for each realization of temperature measurement in time instant $t_i$, are given by,

$$T'(t_i) = T_{obs}(t_i) - T_{pred}(t_i) , \qquad (2)$$

where $T_{pred}$ is the regression equation obtained via that fit. Here $T_{obs}(t_i)$ is the measured temperature.

Figures 11 (a) - (b) depict the fluctuation results for 2 cm above the ground. In Fig. 11 (a), the regression curves are shown, and Fig. 11 (b) presents the temperature residual variation. Fig. 11 (c) gives the histogram of these residuals over the whole series of measurements along with its normal distribution. Table 3 displays variance values for the previously defined time sub-intervals. After 25 min from first contact the inhibition of convection begins to appear falling by approximately one order of magnitude in terms of variance (sub-interval $B_I$). In sub-interval $C_I$, around totality, the variance reaches a minimum (0.004 °C$^2$). At 10:03 the convection begins to increase and, at 10:57 (fourth contact), it is already in a normal level. Note the similar variance values between the extremes sub-interval $A_I$ and $E_I$. The analysis of NSAT fluctuation of residuals indicates that they exhibit almost a normal distribution as exemplified in Fig. 11 (b) or in Fig. 11 (c), with maximum oscillations of $\sim \pm 0.400$ °C in periods of higher convective activity and minimum of $\sim \pm 0.100$ °C near totality. The regression line fitted to all residuals, which is almost exactly equal to zero [Fig. 11 (b)], demonstrates a measure of smoothing quality made at the five different segments or sub-intervals in which the fluctuations were divided.

*5.2 Case II (+ 10 cm)*

On closer inspection, Fig. 4 (or Fig. 9) also reveals appreciable temperature fluctuations during most of the event at +10 cm. Specifically at the eclipse onset, we have sub-interval $A_{II}$ from 08:21 to 08:50 followed by sub-interval $B_{II}$ from 08:51 to 09:11. In the middle, we have sub-interval $C_{II}$ with minimum fluctuations from 09:12 to 10:03. The fluctuations begin to increase again in sub-interval $D_{II}$ from 10:04 to 10:30, and finally they amount to higher values in sub-interval $E_{II}$ from 10:31 to 10:57. The results are shown in Fig. 12 (a) - (b) and in Table 4. The regression curves are portrayed in Fig. 12 (a) and Fig.

12 (b) provides graphically the corresponding residuals. From Table 4 we observe that 29 min past the first contact, the variance is reduced by a factor of 4.3, but in the next interval $C_{II}$, around totality, there is a reduction in the fluctuations by one order of magnitude. In the subsequent two sub-intervals $D_{II}$ and $E_{II}$, the convective activity returns to similar levels as in the first two sub-intervals, respectively.

In comparing sub-interval $C_I$ (Table 3) with sub-interval $C_{II}$ (Table 4), we see that there has been a decline of variance by a factor of 0.5 around and during totality. we see a simultaneous diminishing of convection with height. Graphically, this effect can be noticed examining Fig. 12 (b) in which the minimum fluctuations vary within the range of $\sim \pm 0.050$ °C. Extreme fluctuations of $\sim + 0.400$ °C and $- 0.600$ °C, and of $\sim \pm 0.400$ °C are observed in sub-intervals $A_{II}$ and $D_{II}$, respectively; its corresponding variances in relation to sub-intervals $A_I$ and $D_I$ of Table 3 are quite similar in magnitude. This latter comparison means that no change in convectional activity was detected with a small change of height. Figure 12 (b) also includes the regression line fitted to residuals, which, as before, is almost exactly equal to zero, demonstrating the quality of the smoothing [the histogram and its normal curve of Fig. 12 (c), show evidence of a normal distribution of NSAT residuals].

*5.3 Case III (+ 2 m)*

With regard to the third case, Fig. 4 (or Fig. 10) allows a general picture of the temperature fluctuations at 2 m above the ground. The distribution of these fluctuations with time appears quite different in relation to case II and I. For example, they are appreciable for a longer time (48 min), from first contact (08:21), than the other cases (29 min and 25 min, respectively). This time corresponds to sub-interval $A_{III}$ from 08:21 to 09:09. Next, fluctuations drop off substantially from 09:10 to 09:43 (sub-interval $B_{III}$) approaching totality as expected. Yet, unlike case II and I, fluctuations unexpectedly begin to appear just after the minimum temperature from 09:44 to 10:15 (sub-interval $C_{III}$).

Perhaps other factors or mechanisms, different from convection, were at work, which may explain these specific fluctuations. Afterwards, fluctuations drop off again but in a lesser degree from 10:16 to 10:39 (sub-interval $D_{III}$), and finally, from 10:40 to 10:57 (sub-interval $E_{III}$), convective activity recovers its expected level. In terms of variance (see Table 5), one can realize that the minimum activity, noticed here in sub-interval $B_{III}$, falls even more when compares to case II (sub-interval $C_{II}$) and case I (sub-intervals $B_I$ and $C_I$). This result suggests that the convection suppression due to the eclipse increases with height during the event. Variance in extreme sub-intervals $A_{III}$ and $E_{III}$ display unsurprisingly maximum activity. Continuous sub-intervals $C_{III}$ and $D_{III}$ have similar intermediate values.

A graphic representation of fluctuation variation with time is given in Fig. 13 (b) where the reduction of fluctuations in the middle of the graph is quite clear. The dashed horizontal line in it, which is a regression linear fit, proves the smoothness and quality with which the procedure analysis has been made; Fig. 13 (c) provides the resultant histogram, showing an almost normal distribution of fluctuations. Its normal curve is also shown.

## 6. Discussion

In China other NSAT measurements were made within the TSE shadow track by different teams (Lu et al., 2011; Pintér et al., 2010; Stoeva et al., 2009; Wu et al., 2011; Zainuddin et al., 2001). To some extent a direct or strict comparison of the NSAT changes observed by these teams were not possible given the different ways how they found these changes or due to insufficient information available

Lu's team observed the eclipse from a site located in Chongqing along the Changjiang river; they measured solar radiation, air temperature and *RH* at 1.5 m height at Chongqing University, 245 m above sea level (Huxi campus). In particular, they found that at 8:07, while the eclipse was still in progress, the temperature was still increasing, reaching a maximum of 31.2 ºC at 8:17 before gradually decreasing and reaching a minimum of 28.8 ºC at 9:16; this value was sustained until 9:25 when the temperature began to gradually increase again. Therefore, the NSAT decrease (in this case $\Delta T_{ABS}$) was 2.4 ºC during the eclipse. In order to determine the maximum drop in temperature caused by an eclipse, these authors chose two consecutive days with clear weather for comparison. From their figure 3, it can be inferred that between July 21 and July 22, the variation and value of air temperature were similar outside the eclipse's time interval, and the difference in air temperature across two days was maximized at 9:24, with a maximum difference of 4.6 ºC; therefore, this result gives an idea of how $\Delta T_{LIN}$ was for this eclipse in that location.

In the case of Pintér et al. (2010), whose team observed the eclipse in a site 130 km south-west from Shanghai, their results are presented in their table 2 according to which they made measurements at four heights, namely, 10 cm, 50 cm, 1.5 m and 2.0 m. A drop of 8.0161 ºC, 6.0852 ºC, 4.9582 ºC and 4.4022 ºC were observed, respectively, taken from 1[st] contact to the instant of minimum temperature, which not correspond nor a linear anomaly neither an absolute anomaly as they are defined by Clark (2016). Instead, they correspond to $\Delta T$ as it is defined here in Table 2. Only two of these results are comparable as they correspond to the same height, say, that at 10 cm and 2.0 m; however, the degree to which land or surface type influences the immediate meteorological response to the reduction in solar radiation matters. The ability to compare these results with one another or with other measurements is severely limited because of the possible uncontrolled influences of the surface microclimate on the measurements in each case. This team reported windy conditions, that did not apply to our case, and some clouds during the eclipse. Our instruments were installed in a green-vegetation environment at 809 m above sea level. The vegetative and soil moisture differences of different sites may alter significantly the local response to a radiative reduction by action of a TSE. But, unfortunately, these researchers did not identify the particular micrometeorological environment in which the measurements were taken. The only information that we were able to get from their observation-site coordinates was that they were next to the sea. Thus, any possible comparison is approximate and not identical.

Stoeva's team (Stoeva et al., 2009; Stoev et al., 2012), observed very close to our observation site at three different heights of 10 cm, 50 cm and 2 m; their results are reproduced in Fig. 14 and Table 6 (numerical values of temperature drops are estimated by the difference between the minimum and maximum temperatures plotted in Fig. 14). In comparison to Stoeva et al. (2009) our observed temperature trends are similar, with clouds producing similar effects in the early stages of the eclipse; however, we measure a smaller temperature drop at 10 cm. The difference may be explained by local micrometeorological and observing conditions. They state an instrument resolution of 0.1 °C (manufactured by ExBit, Stara Zagora, Bulgaria) and show images of the equipment setup and its environment. We note that the ground appeared to be a combination of pieces of concrete surrounded by dried greenish-grass. Although our sensors were located very close to where their sensors were, there was a marked differentiation in surface albedo, which could potentially explain some difference between their observations and our measurements.

Pintér et al. (2010), on the other hand, observing from a point close to Shanghai found the same effect as ours in their measurements at four different heights (see Table 6) before totality (see their Figs. 2b, c, d, e), which may also be attributable to clouds at their observing site.

Searching for the effects of this TSE on photo-oxidants in different areas of China, Wu et al. (2011) made observations of NSAT within the umbra at a height of 2 m in Chongqing and Wuhan airports (416 m above sea level and 35 m above sea level, respectively). According to the information given by these authors, taken from a satellite image (see their Fig. 1a), Wuhan was influenced by a light cloud cover but Chongqing was influenced to a lesser extent, and there was no rain during the solar eclipse period according to the local observations in both sites. These observations were modeled and were compared with a simulated run with no yielding, at the instant of minimum temperature, a (quasi-linear) anomaly of 2.12 ºC for the first and 2.03 ºC for the second (see their figs. 3c and 3d). In Table 6 we attempt to give a first comparison of all these results.

As can be observed from this table, based on a comparison between teams, that there is a general decrease tendency of $\Delta T$, $\Delta T_{LIN}$ and $\Delta T_{ABS}$ with height for all places [something similar found by Kapoor et al. (1982) for $\Delta T_{LIN}$ during the 1980 TSE in India]. Also, it seems apparently that there is an approximation between values of $\Delta T_{LIN}$ and $\Delta T_{ABS}$ as long as height increases. The best comparison to be made with our results is that of Stoeva et al.'s results given that both teams were practically at the same observation site.

The team of Zainuddin et al. (2013) observed the eclipse under cloudy conditions at Wild Resort area in Hangzhou (Zhejiang). Their NSAT are shown in their graph 2 where a value of 29.9 ºC is given as maximum temperature at 2nd contact, and a value of 27.9 ºC is given as minimum temperature but before 3rd contact. Thus, a decrease of 2.0 ºC occurred in this interval, which does not correspond to any anomaly considered in this work. Also, these authors do not give at what height above the ground these measurements were taken; hence, this additional anomaly was not included in Table 6.

The activity of the atmospheric boundary layer, extensively studied in the past during this type of phenomenon, were investigated this time, for the 2009 TSE in China, by He et al. (2010) and Chen et al. (2011) from ceilometer observations and meteorological measurements, respectively, over Heifei, not very far away from Shanghai. Yet in our case this activity was studied at Tianhuangping by simply analyzing the fluctuations of air temperature following the same methodology as that we used in our previous paper (Peñaloza-Murillo & Pasachoff, 2015), based on an earlier paper by Szałowski (2002).

In consideration of Tables 3 - 5, giving the variance as indicator of the activity of the atmospheric boundary layer, for different time intervals at different heights, we have the following comparison shown in Table 6. Reading this table horizontally (that is, vs. time) one can see that minimum fluctuations tend to occur at intermediate sub-intervals with emphasis in sub-intervals B and C for all heights; reading it vertically (that is, vs. height) the fluctuations are seldom variable. This result is a clear indication that convective turbulence tends to diminish during the fall in temperature, a typical result analogous to that during daily sunsets. Possibly cloudiness deepens even more the convective turbulence decrease in comparison to a cloudless sky (see Gorchakov et al., 2008; Peñaloza-Murillo & Pasachoff, 2015). In relation to this, these authors, working with the afternoon TOS of March 29, 2006, over Kislovodsk (Russia) and of June 21, 2001, over Lusaka (Zambia, Africa) made analysis of its convective activity calculating the air temperature variance. In those cases, they found higher results than ours indicating less stability. From these last eclipses it can be seen that the tendency towards a steadily and irreversible decreasing convective activity is remarkable for an afternoon TOS in relation to a morning eclipse where there is a reversal of it afterwards [see Figs. 8 - 10 or Figs. 12 - 13 and Fig. 8 of Szałowski (2002)]. However, in comparing our results (Table 7) with those from the latter author it can also be seen that it was found a higher convective activity for the morning 1999 solar eclipse, which was observed at Szczawnica, Poland [see Table 3 of Szałowski (2002)], than that of the morning 2009 TSE at Tinhuangping, a clear indication of the cloudiness effect over convection.

The discussion of the calibration curves involves a matter over a novel method already developed in our previous paper (Peñaloza-Murillo & Pasachoff, 2015), which was based on an early methodology suggested by Phillips (1969). Bearing in mind the cloudiness, it was a complicated task to derive these curves in comparison with the case where the sky was clear, i.e., in Zambia in 2001 (Peñaloza-Murillo & Pasachoff, 2015). Curves like these have never been published before; however, we think that they are consistent and reliable given their performance in reproducing the instantaneous temperature profiles which will be used in our subsequently analysis of NSAT lag in relation to solar radiation (see Fig. 7).

### 7. Conclusions and final comments

Though totality was visible (Petrov et al., 2010; Pasachoff et al., 2011), variable cloudiness complicated observations and our subsequent interpretation of the thermal response of the atmosphere due to the eclipse. Despite the cloudy conditions (Peñaloza-Murillo & Pasachoff, 2018), the method applied in this investigation turns out to be practical and it is affordable to be repeated and rechecked in future similar conditions. In other words, using a simple set of NSAT measurements obtained during solar eclipses even under adverse weather allows a mathematical analysis of these measurements and its relation to changes and fluctuations of it can still be attempted following the methods presented here (Phillips, 1969; Szałowski, 2002) along with that presented by Peñaloza-Murillo & Pasachoff (2018) in which an analysis of the cloudiness and solar radiation during this eclipse was undertaken.

The anomalous discrepancies displayed in Table 6 are indications of the different ambient conditions in which the respective measurements of NSAT were obtained by different teams in China although some trends were, in principle, detected during this eclipse: it is seen that the variations in temperature (or anomalies) decrease as we go up in the atmosphere (Kapoor et al., 1982). Our team confirmed the detection of cloudiness effects over NSAT measurements during the eclipse in terms of the lag reversals found before and after totality. Moreover, the observations made by another two teams, one of them observing far away from our site, seem to reveal also the same effects.

It seems that the technic of applying a virtual variable, defined and used here and in our previous work (Peñaloza-Murillo & Pasachoff, 2015) as "instantaneous temperature," implicitly suggested by Phillips (1969) via "calibration curves," is a workable and convenient way of studying the temperature response during a TSE.

The measurements confirm that the occurrence of a stable stratification in the low boundary layer can be expected due to a total solar eclipse. The general fall of convection intensity is visible in temperature fluctuation component variance decrease. The temporal scales of convection analyzed on the basis of temperature fluctuating component variance, obtained inside 157.2 min (2.62 h) maximum eclipse-centered time range, seems to be different from ones observed under typical or normal conditions of cloudless skies (Peñaloza-Murillo & Pasachoff, 2015). Therefore, it is important to separate the effects of the eclipse from those of the changeable cloud cover, but the increase of the temporal scale seems to take place. The separation of cloudiness effects as well as the issue related to negative lags attributable to cloudiness that we found in this investigation will be a focus of our forthcoming final study on the thermal response of the air temperature near the surface, produced by the longest total solar eclipse of the 21[st] century at Tianhuangping (Zhejiang), China on 22 July 2009, under cloudy skies.

*Acknowledgments*

The participation of the first co-author, MP-M, in the Williams College Expedition to China was possible by a generous partial grant from the Association of Professors of the University of the Andes, APULA (Mérida, Venezuela). The mathematical analysis of the temperature measurements, made during this expedition, were carried out by MP-M as a Fulbright visiting scholar on sabbatical leave (2012), supported by a Fulbright fellowship (Grantee ID: 68110145), at the Hopkins Observatory of Williams College (Williamstown, MA), which he thanks for its hospitality; the help provided by the University of the Andes (ULA) and the international Fulbright program for exchange of scholars, is acknowledged. MP-M also thanks Denise Buell, the Dean of the Faculty of Williams College, for support for his 2018-2019 visits. Co-author MTR wishes to acknowledge the support of his fellow co-author, JMP, for graciously welcoming and hosting him as part of the Williams College expedition to China. MTR also wishes to acknowledge the support of his then graduate advisor at Cornell University, Prof. Peter Gierasch, who happily permitted MTR to take leave with the intent to make measurements and witness the unique beauty of a total solar eclipse. We express our gratitude to Chenyang Sun, Williams College Class of 2022, for his help in the Chinese translation of one of the papers used in this investigation while M.P.-M. was visiting the Astronomy Department of Williams College this year. The Williams College Expedition to China was sponsored in large part by grant 8436-08 from the Committee for Research and Exploration of the National Geographic Society. JMP thanks John Mulchaey for his sabbatical hospitality at the Carnegie Observatories during the completion of this paper. The authors acknowledge to Phil. Trans. R. Soc. A (Royal Society, London) for providing the use of one of the figures published in this article. The eclipse variables (times of first contact, maximum obscuration, last contact, occultation and umbral velocity) were obtained using script available on the website of Xavier M. Jubier [http://xjubier.free.fr]. Data analysis in this study was performed using Microsoft Excel worksheets. The data used in this study are available from the following Williams College website:

https://unbound.williams.edu/facultypublications/islandora/object/facultyarticles%3A188

**TABLE 1**. Astronomical and geographical circumstances for the TSE on 22 July 2009, at Tianhuangping (Mt. Anji), Zhejiang, China. The contacts and other instants are given in local time (LT = UTC + 8 h).

==============================================================

| Circumstances | Value |
| --- | --- |
| Longitude | 119° 35' 28" E |
| Latitude | 30° 28' 07" N |
| Altitude | 809 m |
| Sunrise | 05:12:53 |
| 1st Contact | 08:20:49.1 (8.35 am) |
| Sun altitude | 38.2° |
| 2nd Contact | 09:33:02.8 (9.55 am) |
| Sun altitude | 53.7° |
| Mid-totality | 09:35:53.3 (9.60 am) |
| Sun altitude | 54.3° |
| Minimum apparent distance between lunar center and solar center (at totality) | $6.963 \times 10^{-5}$ rad |
| Occultation | 100% |
| 3rd Contact | 09:38:44.9 (9.65 am) |
| Sun altitude | 55.0° |
| 4th Contact | 10:57:59.5 (10.97 am) |
| Sun altitude | 71.2° |
| Duration of totality | 5' 42.2" (~ 5.70 min) |
| Total duration of eclipse. | 2.62 h (~ 157.2 min) |
| Magnitude at mid-totality* | 1.03071 |
| Solar disk (semi-diameter) | 15' 44.50" ($4.7507 \times 10^{-3}$ rad)[E&A] |
| Lunar disk (semi-diameter) | 16' 19.9" ($4.990 \times 10^{-3}$ rad) |
| Moon/Sun size ratio | 1.10328 |
| Umbral depth | 80.4% |
| Location into the shadow | ~ 30.88 km from the central line[E & A] |
| Shadow path width | ~ 266.66 km[E & A] |
| " minor axis | ~ 262.5 km[E & A] |
| " major " | ~ 300.0 km[E & A] |
| Umbral velocity | 0.872 km/s[†] |
| Sunset | 19:02:17 |

==============================================================

(*) The magnitude of a TSE is the amount of the solar disk's diameter covered by the lunar disk.
(†) Value calculated by Xavier Juvier (see text).
(E & A): Espenak & Anderson (2008).

**TABLE 2**. Meteorological and photometric circumstances of the sky during the TSE on 22 July 2009, at Tianhuangping (Zhejiang), China. Time is given as local (LT = UTC + 8 h); China has one time zone. Air temperature measurements were taken at three different heights above the ground: 2 cm, 10 cm, and 2 m. Values of relative humidity and illumination (in three different directions) were taken from Stoeva et al. (2009)[§].

======================================================================

| *Meteorological* | Value | *Instant/Time Interval/Duration* |
|---|---|---|
| *Sky* | | |
| Cloudiness | Cloudy and variable[†] | 06:31 – 11:50[†] |

| *Solar Radiation* [W/m$^2$] | | |
|---|---|---|
| At 1$^{st}$ contact, $S_{1c}$ | ~ 620[‡] | 08:20:49.1 (8.35 am) |
| At 4$^{th}$ contact, $S_{4c}$ | ~ 760[‡] | 10:57:59.5 (10.97 am) |

| *Air Temperature* [°C] | | | | |
|---|---|---|---|---|
| At height | + 2 cm | + 10 cm | + 2 m | |
| Around sunrise, $T_1$ | 22.56 | 22.82 | 23.93 (±0.25) | 05:12:06 |
| Maximum, $T_{MAX}$ | 36.66 | 33.78 | 31.81 | 11:05:01/10:49:02/11:40:01 |
| At 1$^{st}$ contact, $T_{1c}$ | 30.57 | 30.82 | 29.34 | 08:21:06 (8.35 h) |
| Maxim. (before totality), $T_{max.}$ | 31.36 | 31.69 | 30.14 | 08:27/08:26/08:26 |
| At 2$^{nd}$ contact, $T_{2c}$ | 25.28 | 23.35 | 25.31 | 09:33:05 (9.55 h) |
| " Mid-totality, $T_{Mid}$ | 24.89 | 25.09 | 25.14 | 09:36:00 (9.60 h) |
| " 3$^{rd}$ contact, $T_{3c}$ | 24.65 | 24.97 | 25.02 | 09:39:02 (9.65 h) |
| Minimum (after totality), $T_{min}$ | 24.56 | 24.92 | 24.97 | 09:41:01/09:40:05/09:40:05 |
| At 4$^{th}$ contact, $T_{4c}$ | 35.31 | 33.16 | 31.13 | 10:58:01 (10.97 h) |
| At last measurement, $T_f$ | 28.27 | 27.04 | 25.23 | 14:54:02 (14.90 h) |
| $\Delta T = T_{1c} - T_{2c}$ (reduction) | 5.29 | 5.47 | 4.03 (±0.25) | ~ 71.98' (~ 1.2 h) |
| $\Delta T$" = $T_{2c} - T_{min}$ ( " ) | 0.73 | 0.44 | 0.34 | [~7.98'/~7.08'/~7.08']* |
| $\Delta T = \Delta T' + \Delta T$" (reduction) | 6.01 | 5.90 | 4.37 | ~80'(1.3h)/~79'(1.32h)/~79'(1.32h) |
| $T_{i,n-e}$ (interpolated,non-eclipse) | 32.98 | 31.99 | 30.24 | 09:41:01/09:40:05/09:40:05 |
| $\Delta T_{LIN} = T_{h,n-e} - T_{min}$ | 8.42 | 7.07 | 5.27 | 09:41:01/09:40:05/09:40:05 |
| $\Delta T_{ABS} = T_{max} - T_{min}$ | 6.80 | 6.77 | 5.17 | 80.01'/74.28'/74.28' |

| *Relative Humidity* [%] | | |
|---|---|---|
| At first measurement, $RH_1$ | 70 | 08:00:00 |
| Minimum, $RH_{min}$ | 60 | 08:43:00 – 09:10:00 |
| Maximum, $RH_{max}$ | 72 | 09:45:00 |
| At last measurement, $RH_f$ | 59 | 10:27:00 |
| $\Delta RH' = RH_1 - RH_{min}$ | 10 | 43' |
| $\Delta RH$" = $RH_{max} - RH_{min}$ | 12 | 35' |

| *Wind speed* | - | - |
|---|---|---|

| *Photometric* | Value | *Instant* |
|---|---|---|

| Ilumination [lx] | Zenith | Horizon | Solar Area | |
|---|---|---|---|---|
| At first measurement, $I_1$ | 274 | 48 | 205 | 07:08:36 |
| Minimum, $I_{min}$ | 17 | 4 | 120 | 09:34:50 |
| At last measurement, $I_f$ | 3095 | 518 | 3287 | 11:27:18 |

========================================================

(§) For similar and different measurements made during this eclipse at Tianhuangping, and in other sites of China as well as in South Korea and India, see Pintér et al. (2010), He et al. (2010), Chen et al. (2011), Zainuddin et al. (2013), Kwak et al. (2011), Chung et al. (2010), Wu et al. (2011), Lu et al. (2011), Kumar (2014), Rao et al. (2013), Jeon (2011), Wang & Liu (2010) and Jeon & Oh (2011).

(‡) Values derived from calculations [see Fig. 4 and also Figs. 14 and 15 of Peñaloza-Murillo & Pasachoff (2018)].

(†) See Peñaloza-Murillo & Pasachoff (2018).

(*) These are the lags for air temperature reduction $\Delta T''$ at specified heights measured from the time of second contact. For the same spot, but in a different environment where their measurements were taken, Stoeva et al. (2009) observed a lag of ~ 1 min after the end of total phase for three different heights above the ground, namely, of +10 cm, +50 cm and +2 m.

**TABLE 3**. Time sub-intervals for analysis of the NSAT fluctuations, at +2 cm, due to convection during the TSE on 22 July 2009, at Tianhuangping (Zhejiang), China. The table includes the results for the coefficients of determination values and variance per sub-interval.

========================================================

| | | | Time Sub-Intervals | | |
|---|---|---|---|---|---|
| | $A_I$ | $B_I$ | $C_I$ | $D_I$ | $E_I$ |
| Interval | 08:21 - 08:46 | 08:47 - 09:06 | 09:07 - 10:02 | 10:03 - 10:35 | 10:36 - 10:57 |
| Duration (min) | 25 | 19 | 55 | 32 | 21 |
| Coefficient of determination ($R^2$) | 0.8709 | 0.9497 | 0.9979 | 0.0031 | 0.9193 |
| Variance (°C$^2$) | 0.043 | 0.005 | 0.004 | 0.030 | 0.042 |

========================================================

**TABLE 4**. Time sub-intervals for analysis of the NSAT fluctuations, at +10 cm, due to convection during the TSE on 22 July 2009, at Tianhuangping (Zhejiang), China. The table includes the results for the coefficients of determination values and variance per sub-interval.

========================================================

| | | | Time Sub-Intervals | | |
|---|---|---|---|---|---|
| | $A_{II}$ | $B_{II}$ | $C_{II}$ | $D_{II}$ | $E_{II}$ |
| Interval | 08:21 - 08:50 | 08:51 - 09:11 | 09:12 - 10:03 | 10:04 - 10:30 | 10:31 - 10:57 |
| Duration (min) | 29 | 20 | 51 | 26 | 26 |
| Coefficient of determination ($R^2$) | 0.9519 | 0.8032 | 0.9979 | 0.9909 | 0.9269 |

| Variance ($°C^2$) | 0.043 | 0.010 | 0.002 | 0.012 | 0.038 |

**TABLE 5**. Time sub-intervals for analysis of the NSAT fluctuations, at +2 m, due to convection during the TSE on 22 July 2009, at Tianhuangping (Zhejiang), China. The table includes the results for the coefficients of determination values and variance per sub-interval.

| | Time Sub-Intervals | | | | |
|---|---|---|---|---|---|
| | $A_{III}$ | $B_{III}$ | $C_{III}$ | $D_{III}$ | $E_{III}$ |
| Interval | 08:21 - 09:09 | 09:10 - 09:43 | 09:44 - 10:15 | 10:16 - 10:39 | 10:40 - 10:57 |
| Duration (min) | 48 | 33 | 31 | 23 | 17 |
| Coefficient of determination ($R^2$) | 0.9519 | 0.8032 | 0.9979 | 0.9909 | 0.9269 |
| Variance ($°C^2$) | 0.050 | 0.001 | 0.010 | 0.009 | 0.020 |

**TABLE 6**. Different air temperature negative anomalies near the surface produced by the longest total solar eclipse of the 21st century on 22 July 2009, at different heights and sites of China.

| Height [cm] | $\Delta T$ [ºC] | $\Delta T_{LIN}$ [ºC] | $\Delta T_{ABS}$ [ºC] |
|---|---|---|---|
| 2 | **6.01** | **8.42** | **6.80** |
| 10 | **5.90** 8.02[P] | **7.07** | **6.77**, 8.4[S] |
| 50 | 6.01[P] | - | 6.5[S] |
| 150 | 4.96[P] | 4.6[L] | 2.4[L] |
| 200 | **4.37** 4.40[P] | **5.27** [2.12[Ch], 2.03[Wuh]][W] | **5.17**, 4.7[S] |

Bold figures are values of this work.
(P) Values from Pintér et al. (2010) estimated from measurements taken 130 km south from Shanghai at approximately 0 m above sea level (asl).
(S) Values of Stoeva et al. (2009) taken very close to our observation site (809 m asl) by that team.
(W) Values from Wu et al. (2011) taken at Chongqing airport (Ch) (416 m asl) and Wuham airport (Wuh) (35 m asl).
(L) Values from Lu et al. (2011) taken at Huxi campus, Chongqing University (245 m asl).

**TABLE 7**. Air temperature variance at different heights during the TSE on 22 July 2009, at Tianhuangping. The time sub-intervals are not equal but have approximately the same length (see Tables 3 - 5). Near totality the atmospheric layer near the surface seems to become rather stable.

| Height | Variance ($°C^2$) | | | | |
|---|---|---|---|---|---|
| | sub-intervals $A_{I,II,III}$ | sub-intervals $B_{I,II,III}$ | sub-intervals $C_{I,II,III}$ | sub-intervals $D_{I,II,III}$ | sub-intervals $E_{I,II,III}$ |
| +2 cm (I) | 0.043 | 0.005 | 0.004 | 0.030 | 0.042 |
| +10 cm (II) | 0.043 | 0.010 | 0.002 | 0.012 | 0.038 |
| +2 m (III) | 0.050 | 0.001 | 0.010 | 0.009 | 0.020 |

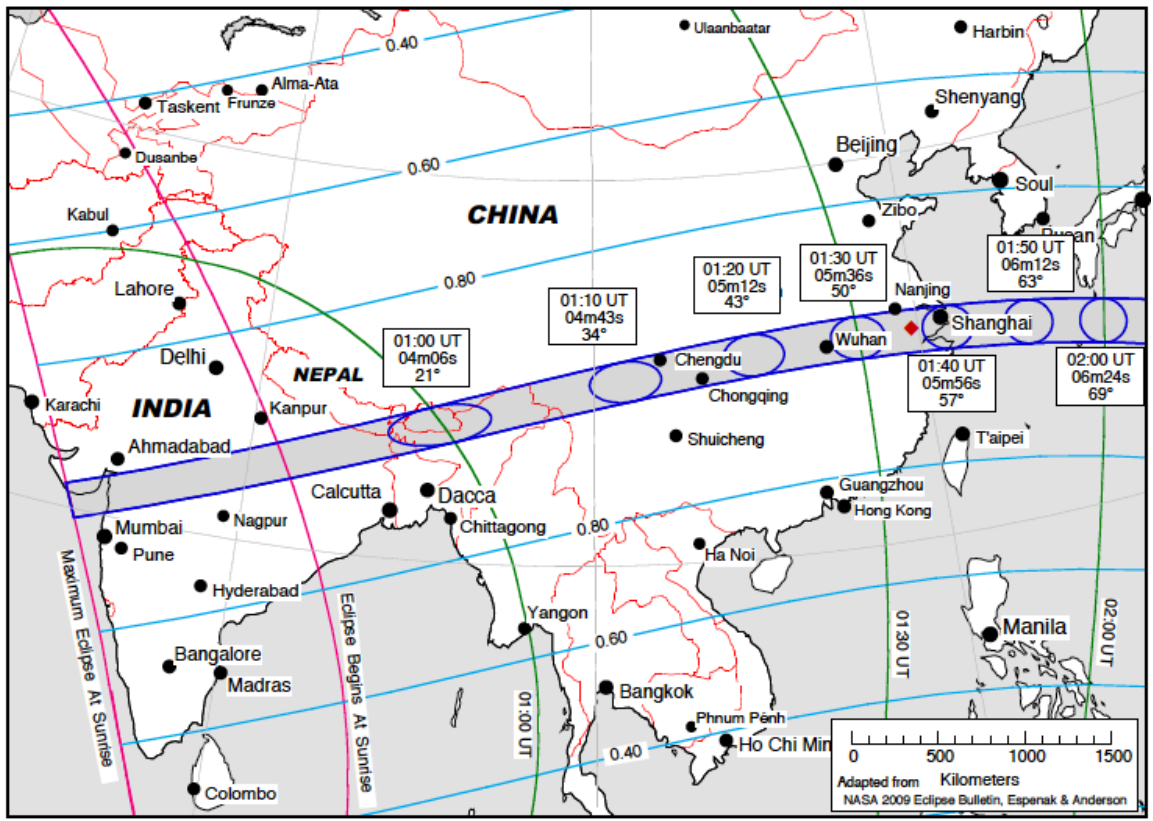

**FIGURE 1**. The umbral and penumbral region of the solar eclipse on 22 July 2009, over Asia. The totality path passed over India, Nepal, Bangladesh, Bhutan, Burma, China, and some Japanese islands (Reprinted from Espenak & Anderson, 2008). Our observation site, in the mountains near Hangzhou west of Shanghai, had nearly the longest possible duration available on a large land mass (see Peñaloza - Murillo & Pasachoff, 2018). The red diamond indicates approximately our observation site.

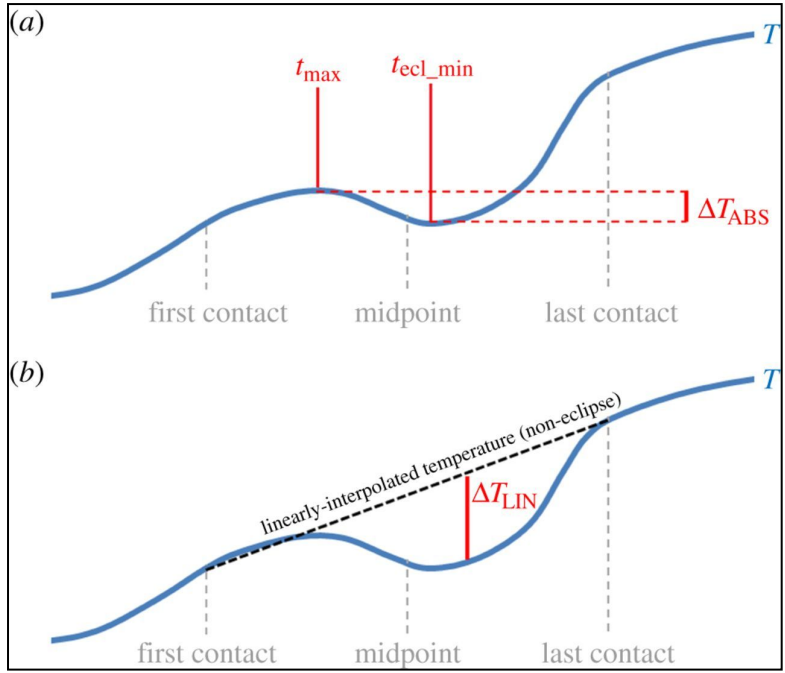

**FIGURE 2**. The graphs shown in this figure, taken from Clarke (2016), illustrate schematically how absolute (a) and linear (b) anomalies are defined. Here $t_{max}$ in our notation is $T_{max}$ and $t_{ecl\_mim}$ is $T_{min}$. Thick blue line denotes the observed temperature. Note that $\Delta T_{LIN} > \Delta T_{ABS}$.

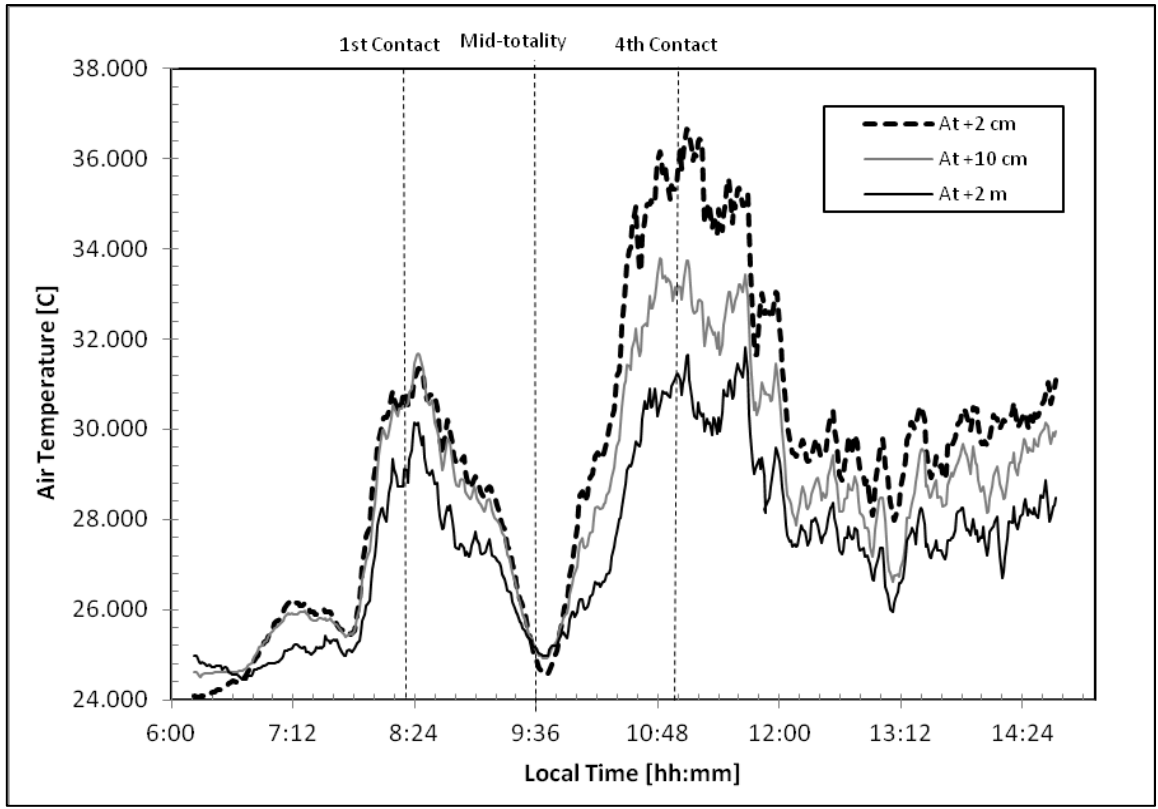

**FIGURE 3**. Air temperature variation [in ºC], at three different heights, during the day of TSE on 22 July 2009, in Tianhuangping (Mt. Anji, Zhejiang), China. The measurements were stopped at around 14:40 due to bad weather (overcast and rain). The effect of the eclipse on measurements is clearly seen as well as the effect of bad weather around 11:30 onwards. These noisy patterns are the result of different combined factors like cloudiness, convection, altitude, solar radiation, etc., that intervene in determining the temperature variation over time. Uncertainties are estimated to be ± 0.25 ºC.

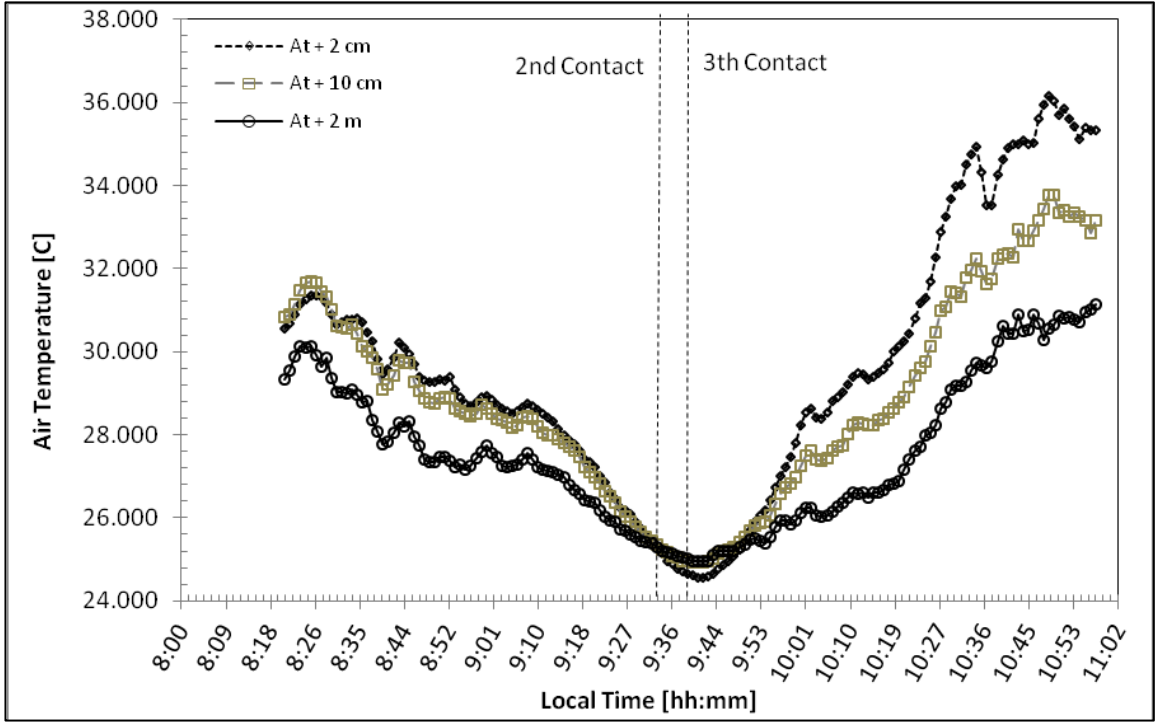

**FIGURE 4**. Air temperature variation [in °C] during the TSE on 22 July 2009, at Tianhuangping at different indicated heights. First contact was at 08:20:49.1 and fourth one at 10:57:59.5. Totality occurred between 09:33:02.8 (2nd contact) and 09:38:44.9 (3th contact).

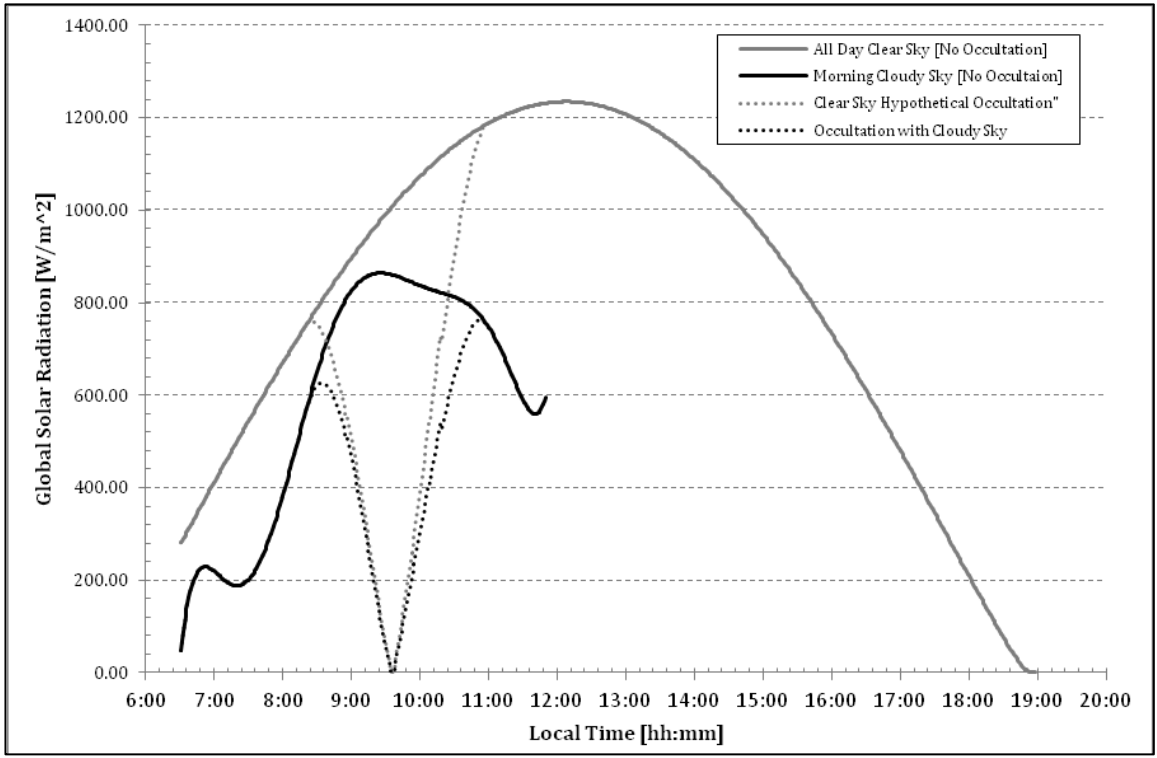

**FIGURE 5**. Theoretical model for global solar radiation [in W/m²] where the eclipse is considered under cloudless (grey lines) and cloudy (black lines) conditions at Tianhuangping during the morning on 22 July 2009, obtained by Peñaloza-Murillo & Pasachoff (2018).

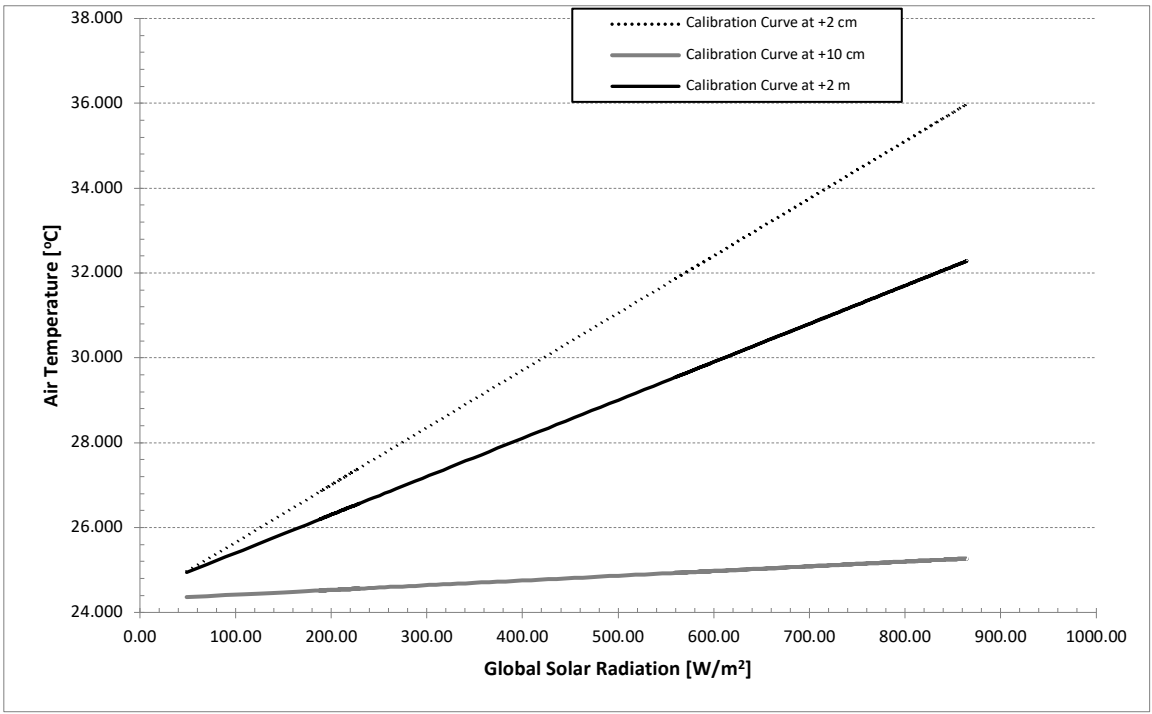

**FIGURE 6**. Approximate calibration curves for the three indicated heights above the ground. These curves enable converting solar flux during the eclipse (Fig. 4) into the so-called instantaneous temperature (Fig. 5). They were estimated following the procedure of Phillips (1969), which was subsequently applied by Peñaloza-Murillo & Pasachoff (2015).

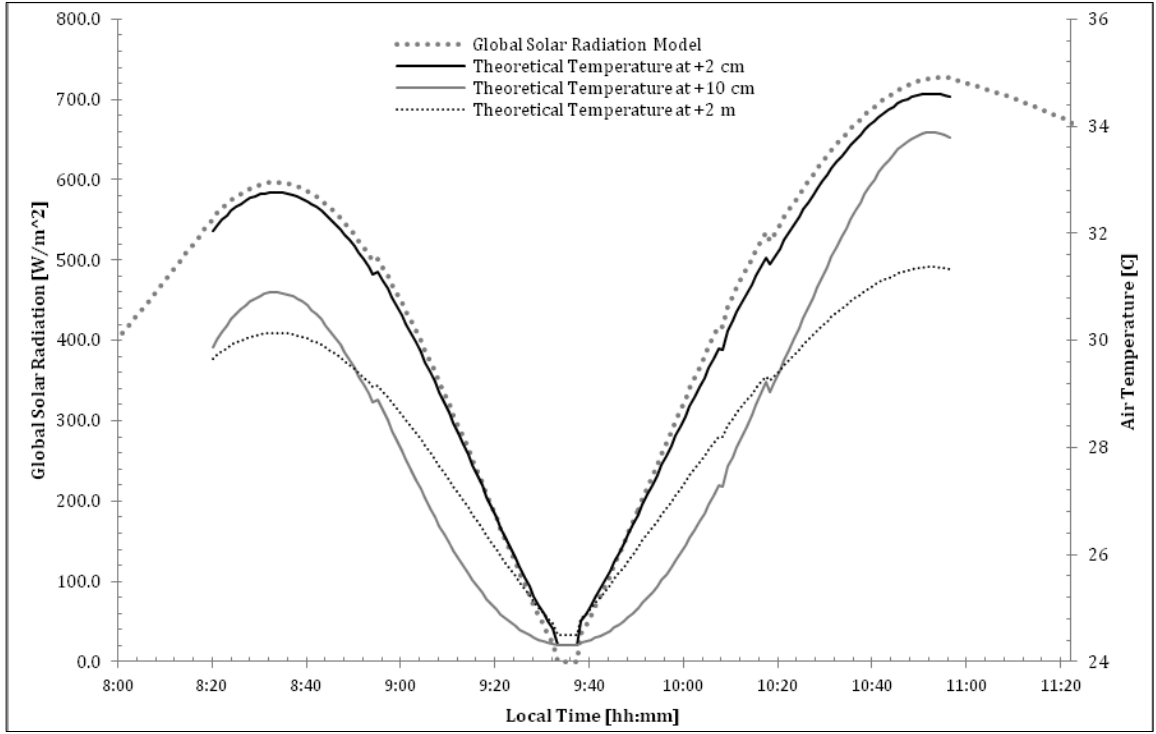

**FIGURE 7**. Theoretical curves illustrating our radiative model for global solar radiation (grey dots) [in W/m²] and instantaneous air temperature [in ºC], at the indicated heights, as a result of the TSE on 22 July 2009, in Tianhuangping (Mt. Anji, Zhejiang, China). Because they are all in phase, the models for this instantaneous temperature are used, instead of the global solar radiation variation due to the eclipse to study the NSAT lag and the delay function (the kinks of the curves observed at 8:50, 10:10 and 10:20 are artificial and come out from calculations of the occultation function).

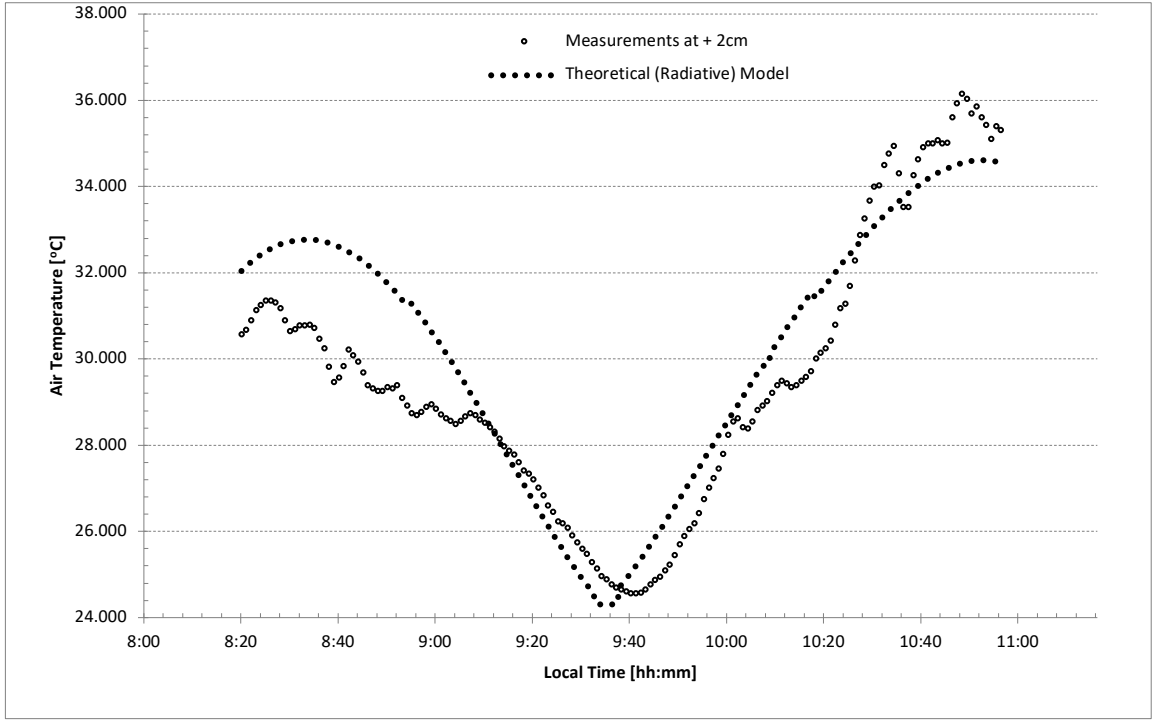

**FIGURE 8**. Theoretical radiative model (black dots) of NSAT at 2 cm above the ground. This theoretical curve is considered to be the virtually instantaneous response of the air to the TSE on 22 July 2009, at that height in Tianhuangping, because it is in phase with the solar radiation model described in Figs. 5 and 7. Note that before totality approximately between 09:04 and 09:12 measured temperature coincides with $T_{inst}$, having no lag in that interval. In the previous interval there is a clear and significative unexpected negative lag until 09:12. After totality there are a least three times where both temperatures are equal, at 10:02, 10:28 and 10:36, approximately. Yet from the second of these times onwards there is an unexpected reversal of the lag with a small fluctuation where this third time is.

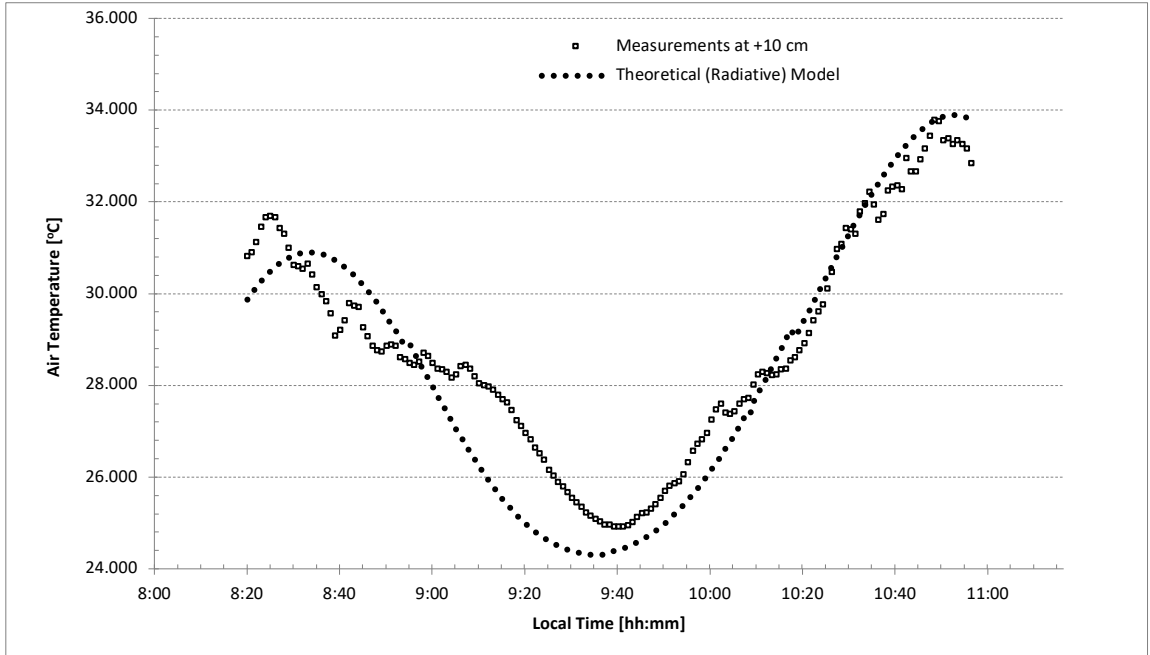

**FIGURE 9**. Theoretical radiative model (black dots) of NSAT at 10 cm above the ground. Similarly, it is considered as the virtually instantaneous response of the air to the TSE on 22 July 2009, at that height in Tianhuangping, because it is in phase with the solar radiation model described in Figs. 5 and 7. Note that before totality there are at least two times where measured temperature is equal to $T_{inst}$, approximately at 08:30 and 08:57; in between, there is an unexpected negative lag. After totality, at 10:13 measured temperature is equal to $T_{inst}$. From this time up to 10:25 there is an unexpected reversal of the lag. From this time up to 10:45, approximately, values of measured temperature and $T_{inst}$ are practically the same; after that this lag reversal continues until the end of the eclipse.

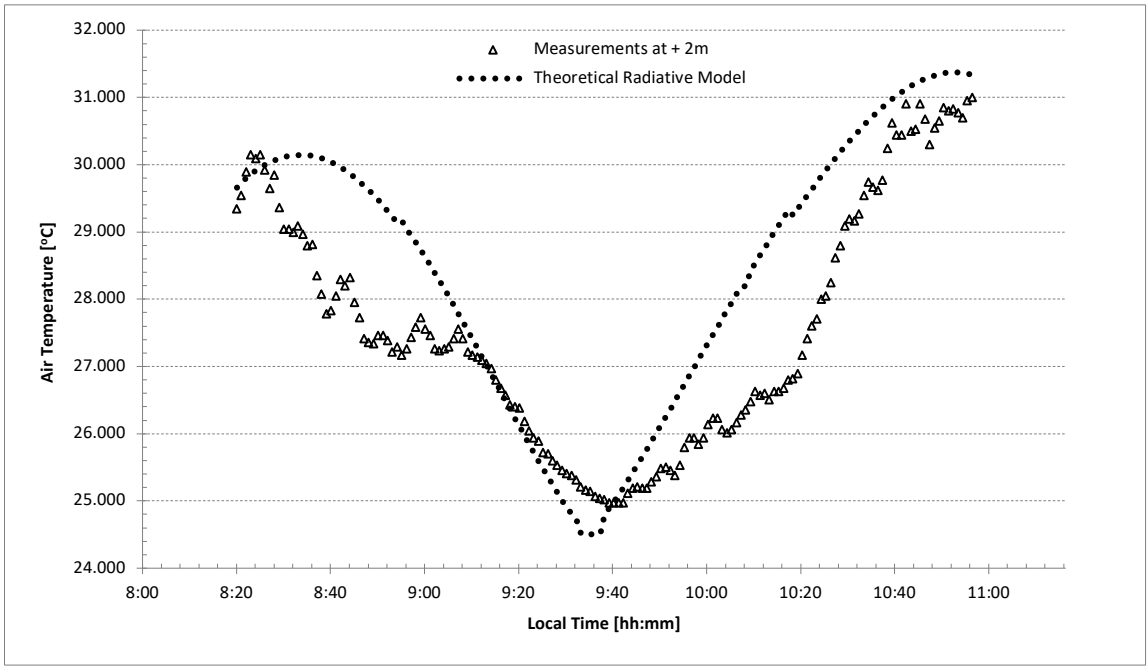

**FIGURE 10**. Theoretical radiative model (black dots) of NSAT at 2 m above the ground. Equally, it is considered as the virtually instantaneous response of the air to the TSE on 22 July 2009, at that height in Tianhuangping, because it is in phase with the solar radiation model described in Figs. 5 and 7. As in the previous cases, there is an unexpected negative lag before totality, between 08:26 and 09:12. Between 09:12 and 09:19 values of measured temperature are equal to $T_{inst}$. Afterward, the lag is as expected for the rest of the eclipse.

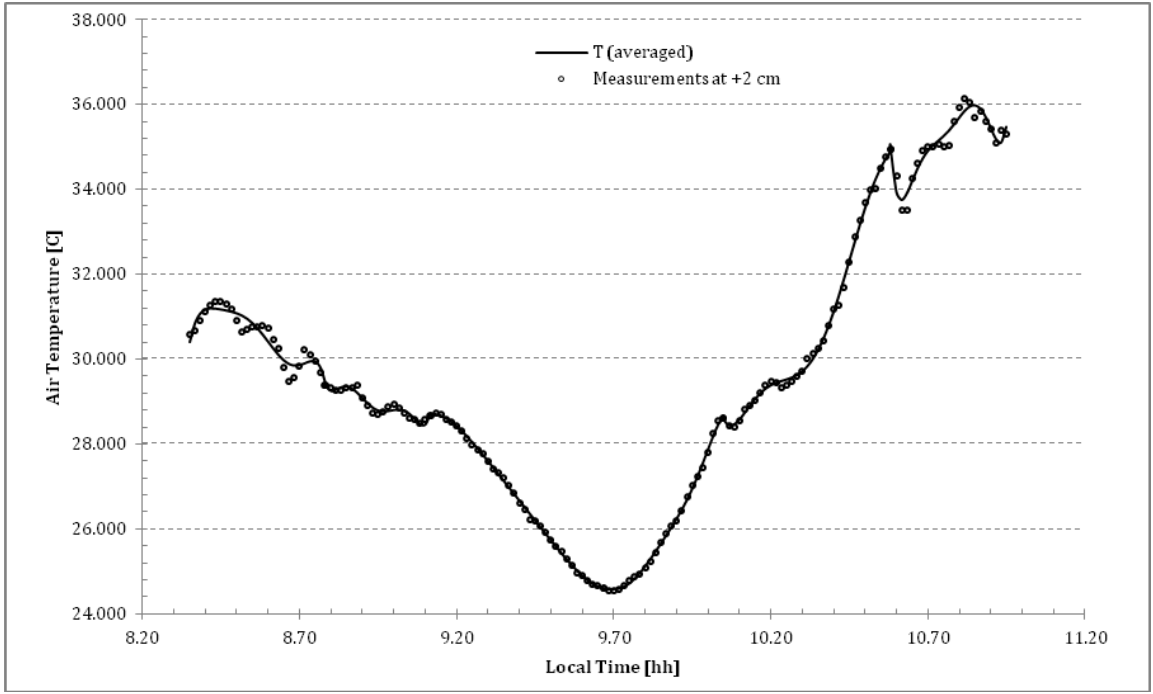

**FIGURE 11 (a)**. The model of mean temperature (solid black line) [in ºC] fit to fluctuations in the measurements made at 2 cm above the ground. A polynomial regression fit has been applied sector-by-sector depending on the dispersion of measurements.

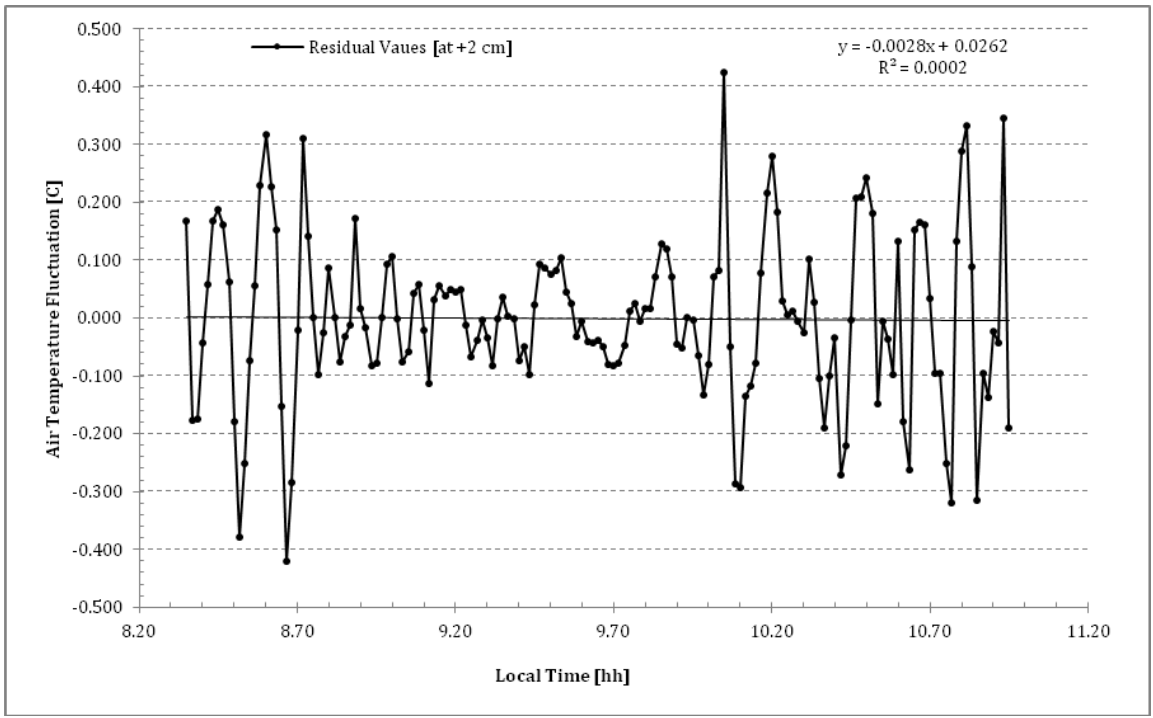

**FIGURE 11 (b)**. NSAT residuals around the mean temperature [in ºC] given by regression fit in Fig. 11 (a), at +2 cm. It is clearly seen a decreasing of temperature fluctuations between 8.80 h (09:48) and 10.00 h (10:00) due to the eclipse; outside of this interval, the fluctuations are greater. The regression line fitted to these residuals (almost zero) indicates the quality of the smoothing procedure.

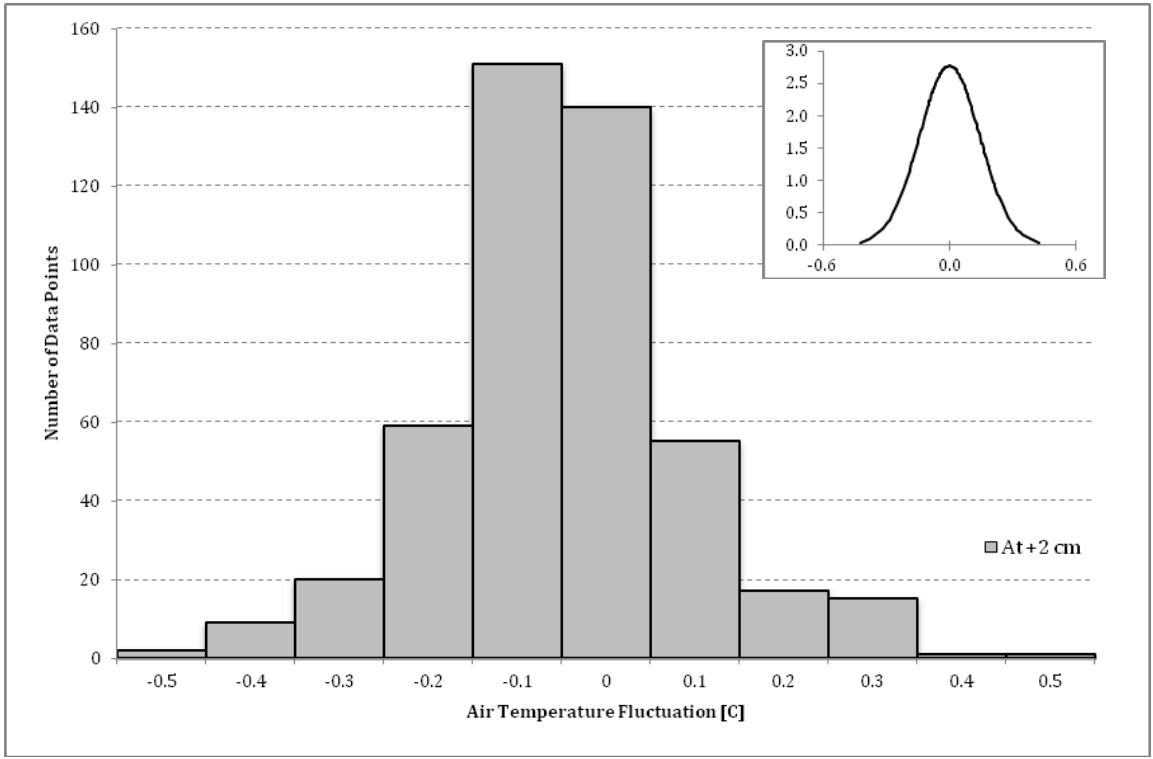

**FIGURE 11 (c)**. The histogram of NSAT residuals [in ºC], over the whole series of measurements, presented in Fig. 11 (b) at 2 cm above the ground. A normal distribution curve fit is shown in the upper corner inset.

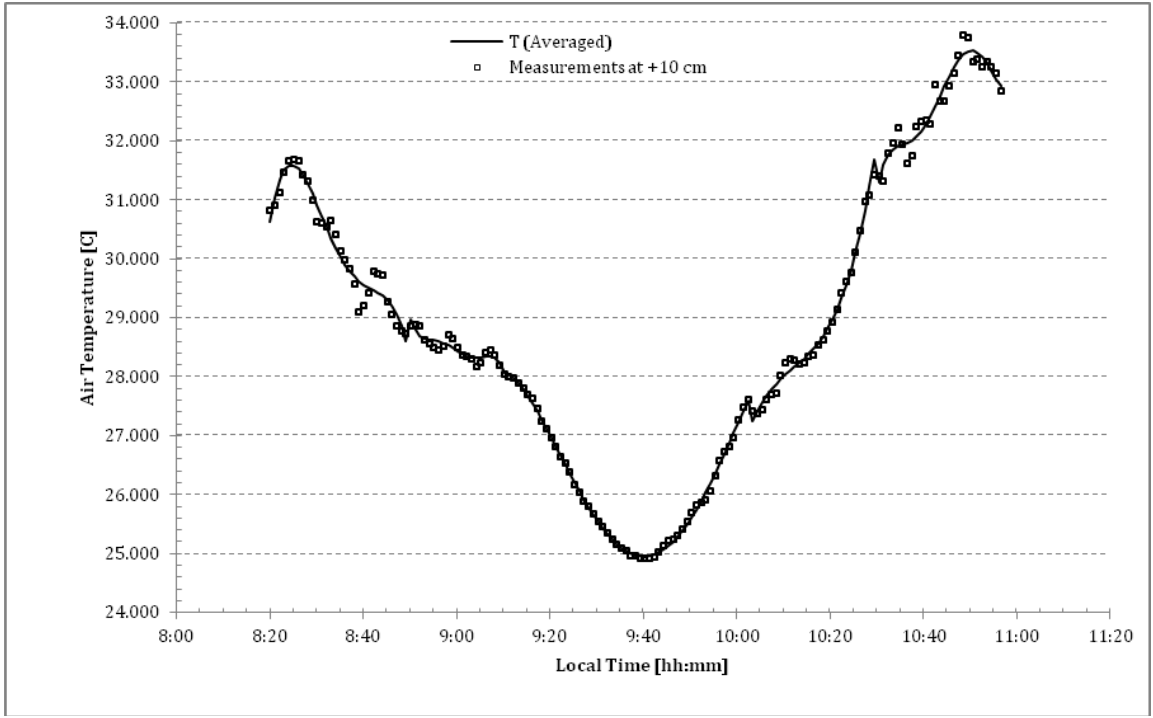

**FIGURE 12 (a)**. The model of mean temperature (solid black line) [in ºC] fit to fluctuations in the measurements made at 10 cm above the ground. A polynomial regression fit has been applied sector-by-sector depending on the dispersion of measurements.

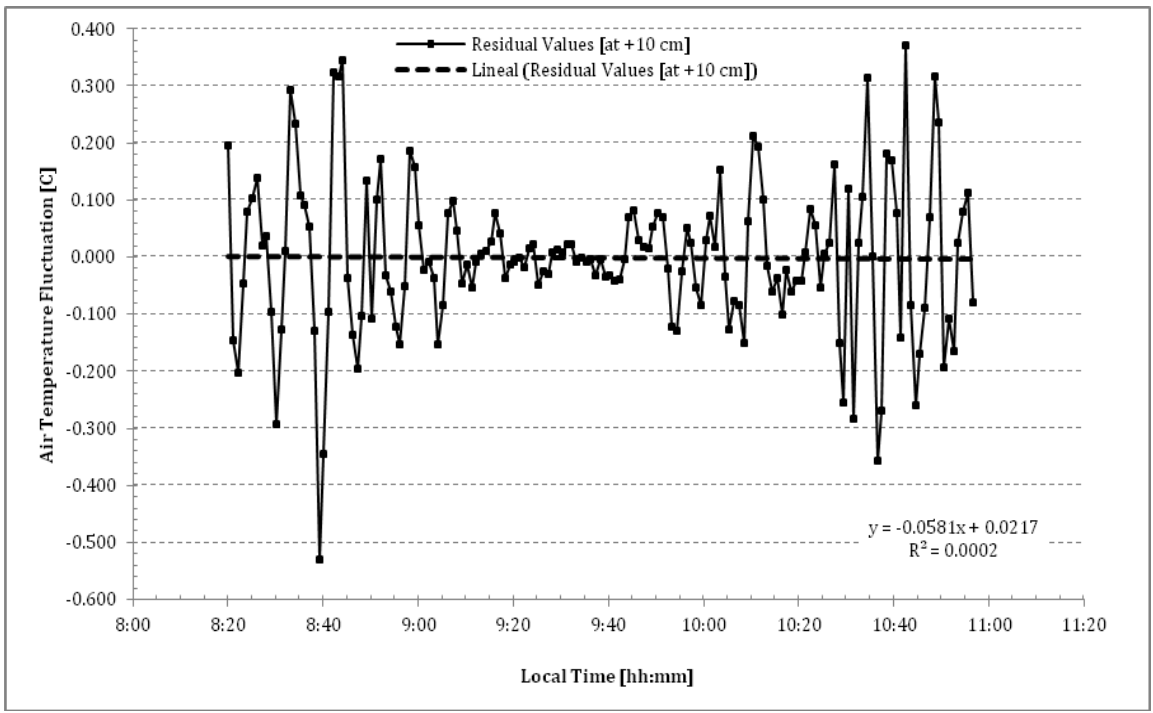

**FIGURE 12 (b)**. NSAT residuals [in ºC] around the mean temperature given by a regression fit in Fig. 12 (a), at +10 cm. A reduction of temperature fluctuations even greater than at +2 cm, between 09:10 and 09:45 due to the eclipse is noted; out of this interval, the fluctuations are greater. The regression (dashed) line fitted to these residuals (almost zero) indicates the quality of the smoothing procedure made through the entire period of time.

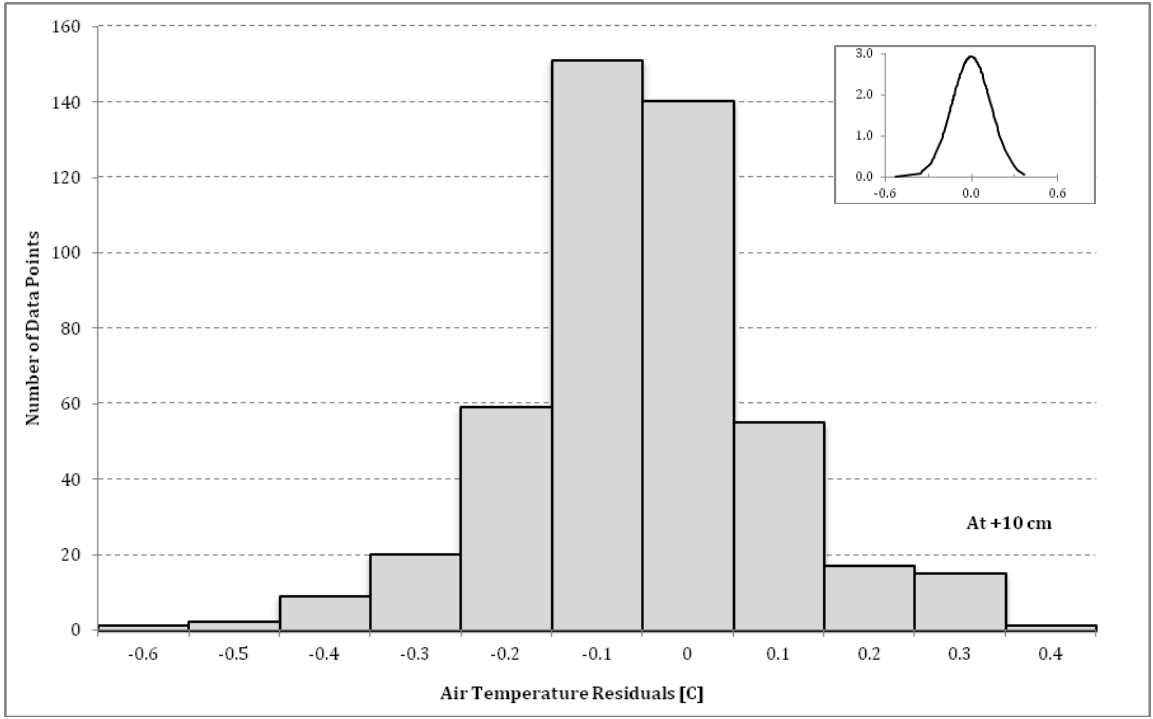

**FIGURE 12 (c)**. The histogram of air temperature residuals [in ℃], over the whole series of measurements, presented in Fig. 12 (b) at 10 cm above the ground. A normal distribution curve fit is shown in the upper corner inset. It is skewed a little to the left.

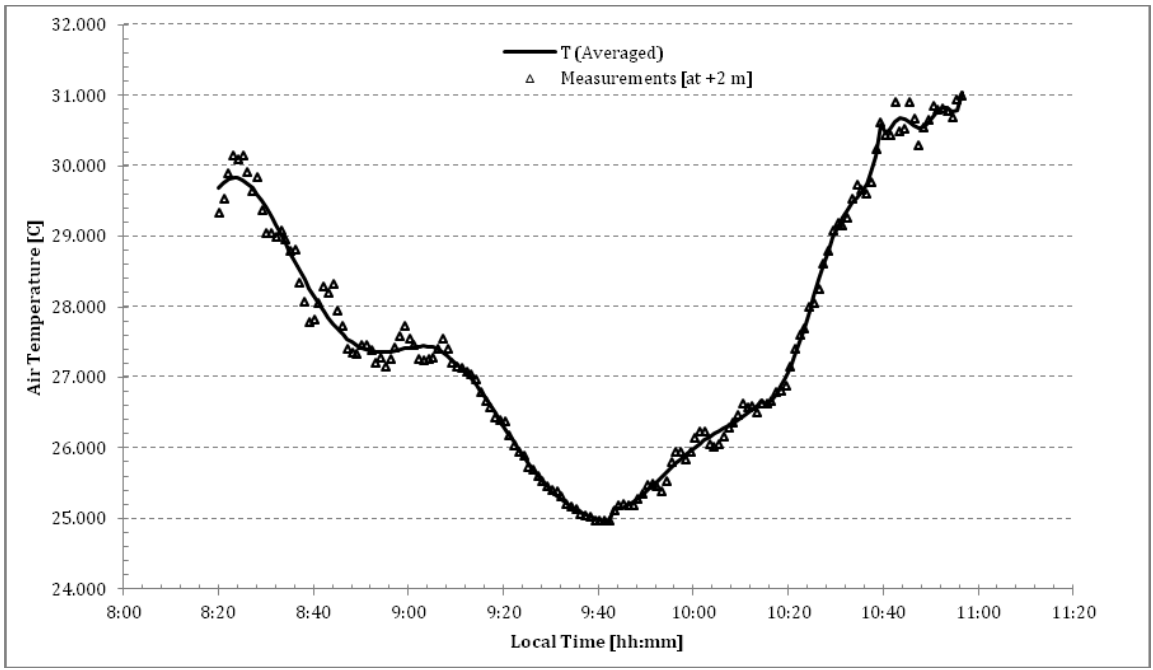

**FIGURE 13 (a)**. The model of mean temperature (solid black line) [in ºC] fit to fluctuations in the measurements made at 2 m above the ground. A polynomial regression fit has been applied sector-by-sector depending on the dispersion of measurements.

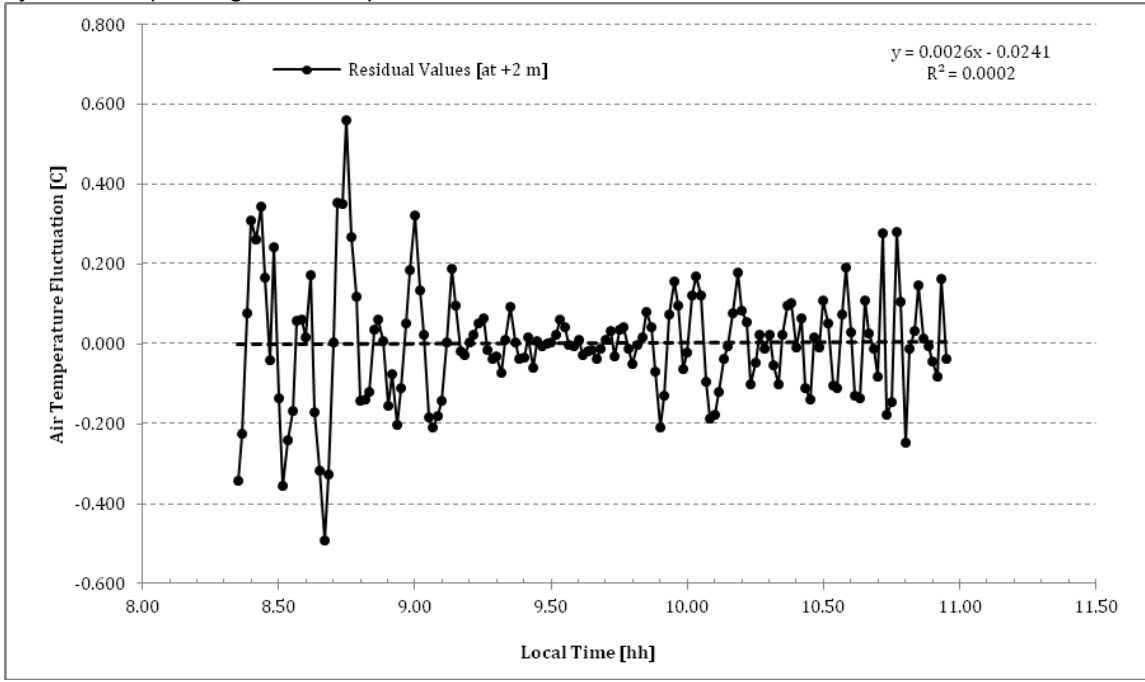

**FIGURE 13 (b)**. NSAT residuals [in ºC] around the mean temperature given by regression fit in Fig. 13 (a), at 2 m. The reduction of temperature fluctuations observed between 9.2 h (09:12) and 9.8 h (09:48) due to the eclipse, is even greater than at +10 cm; out of this interval, the fluctuations are greater. The regression (dashed) line fitted to these residuals (almost zero) indicates the quality of the smoothing procedure.

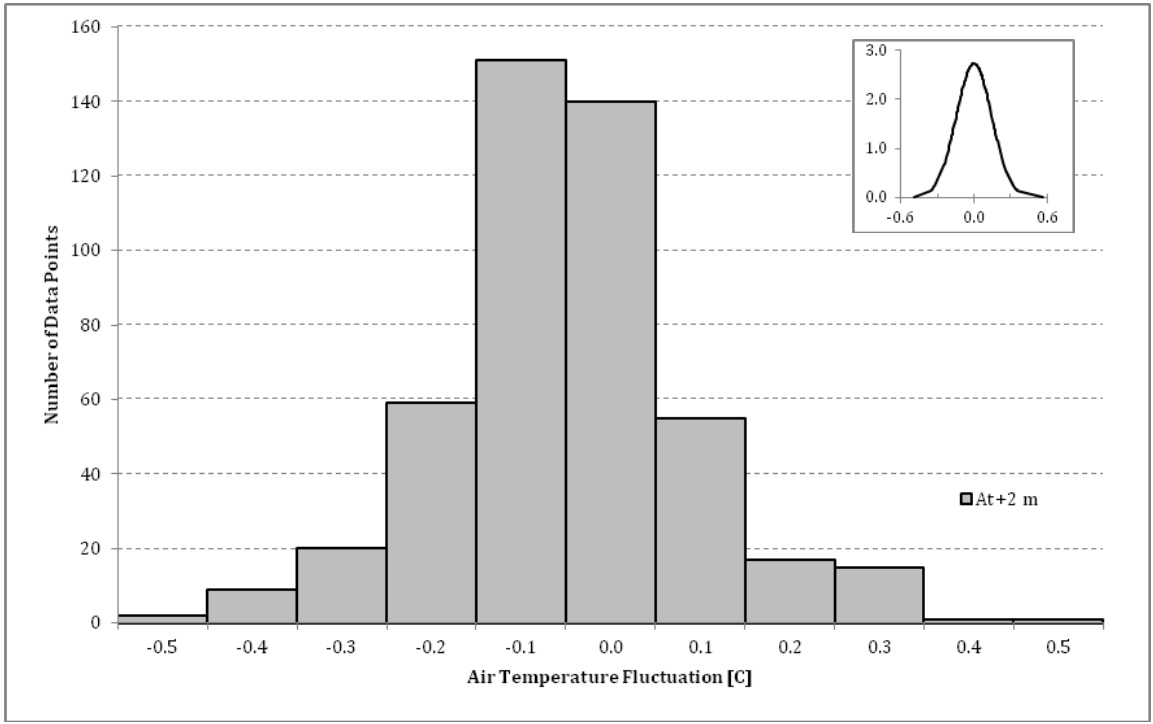

**FIGURE 13 (c)**. The histogram of air temperature residuals [ºC], over the whole series of measurements, presented in Fig. 13 (b) for 2 m above the ground. A normal distribution-curve fit is shown in the upper corner inset. Also, it is skewed a little to the right.

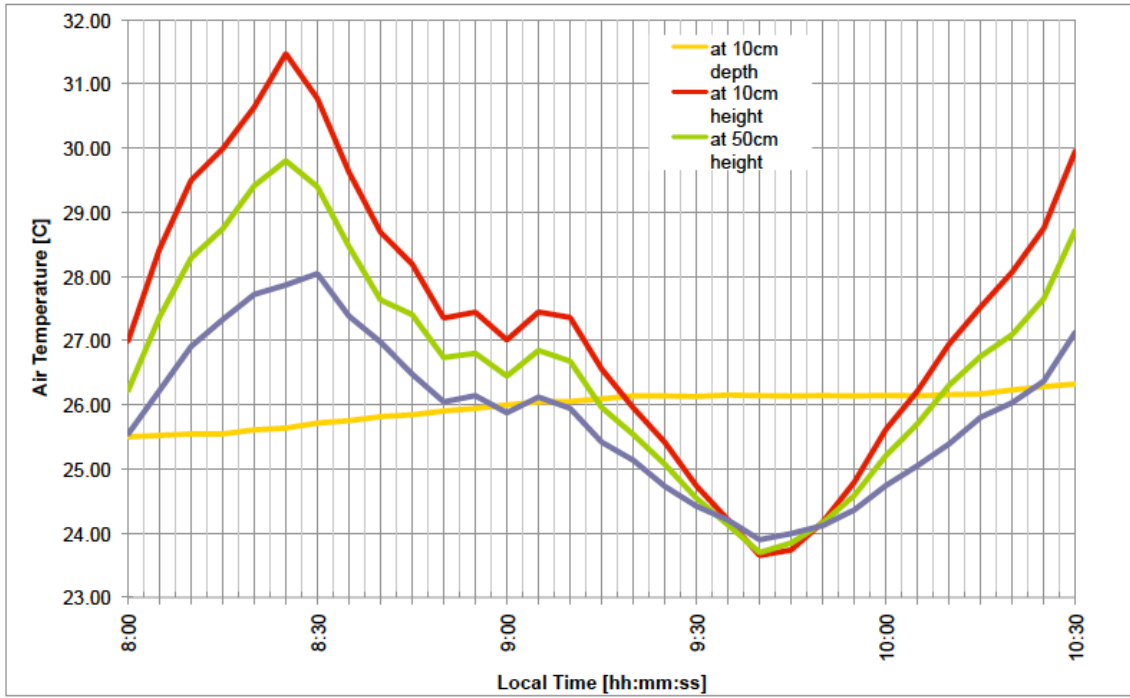

**Figure 14**. NSAT measurements [in ºC] made by Stoeva at al. (2009) at Tianhuangping during the 2009 TSE, between 08:00 am and 10:30 am (LT). The blue line corresponds to a height of 200 cm above the ground.